\documentclass[11pt]{article}

\usepackage[margin=1in]{geometry}
\usepackage{amsmath,amssymb,amsthm,mathtools,booktabs,graphicx,array}
\usepackage[shortlabels]{enumitem}
\usepackage{natbib}
\usepackage[T1]{fontenc}
\usepackage{lmodern}
\usepackage{microtype}
\usepackage[dvipsnames]{xcolor}
\usepackage{hyperref}
\hypersetup{colorlinks=true,linkcolor=blue,citecolor=blue,urlcolor=blue}

\newtheorem{proposition}{Proposition}
\newtheorem{remark}{Remark}

\newcommand{\E}{\mathbb{E}}
\newcommand{\Var}{\operatorname{Var}}
\newcommand{\Normal}{\mathcal{N}}
\newcommand{\GammaDist}{\operatorname{Gamma}}
\newcommand{\given}{\,|\,}
\newcommand{\dd}{\mathrm{d}}
\newcommand{\fcdpglm}{\textit{fc}DPGLM}
\renewcommand{\th}{\theta}

\title{\textbf{A Semiparametric Bayesian Generalized Linear Mixed Model with a Nonparametric Count Distribution for Seizure Data}}
\author{Entejar Alam\thanks{Department of Statistics and Data Sciences, University of Texas at Austin. Email: \texttt{entejar@utexas.edu}}
\and Elizabeth Juarez-Colunga\thanks{Department of Biostatistics and Informatics, University of Colorado Anschutz}
\and Marco I.\ Gonzalez\thanks{Department of Neurology, University of California, Davis}
\and Peter M\"uller\thanks{Department of Statistics and Data Sciences, University of Texas at Austin; Department of Mathematics, University of Texas at Austin}}
\date{}

\begin{document}
\maketitle

\begin{abstract}
Functional count data arise when recurrent events are recorded repeatedly over time and scientific interest centers on how event rates evolve over time and across experimental conditions. Standard mixed effects models for count data include interpretable mean structures but require a parametric conditional distribution, which can be restrictive in the presence of overdispersion, unusual zero frequencies and heavy tails. To address these challenges, we develop the functional count DPGLM (\textit{fc}DPGLM) as a semiparametric Bayesian mixed model that combines GLM-style regression with a nonparametric count distribution. Parametric condition-specific mean trajectories and subject-specific random effects determine the conditional mean exactly, allowing the interpretation of contrasts as usual, while remaining details of the count distribution are learned nonparametrically from the data. Poisson and negative-binomial mixed models are recovered as limiting cases. A gamma completely random measure representation allows for efficient posterior simulation that combines a Fisher-preconditioned Langevin update with an exact update of the total random-measure mass. In simulations that are designed to mimic a rat seizure study, \textit{fc}DPGLM estimated functional mean contrasts as accurately as inference under a correctly specified negative-binomial mixed model, and more accurately otherwise, and recovered probabilities for zero counts and exceedance probabilities that remain misrepresented under the parametric models. We apply the model to inference for seizure data over time for rats. Seizures were recorded in two periods after induced status epilepticus, an early period and a chronic period. Seizure rates rise and then decline in the early period but remain high and nearly flat in the chronic period, and rats differ substantially. Poisson and negative-binomial mixed models estimate similar mean trajectories and period contrasts, but differ in distributional details: \textit{fc}DPGLM matches the observed zero frequency, variance and high-burden days, which the Poisson and negative-binomial models do not.
\end{abstract}

\noindent\textbf{Keywords:} Bayesian nonparametrics; completely random measure; epilepsy; exponential tilting; generalized linear mixed model; overdispersion.

\section{Introduction}
\label{sec:intro}

We introduce a semiparametric Bayesian generalized linear mixed model (GLMM) for
recurrent count data. We build on recently proposed semiparametric generalized
linear models (GLM) to introduce the first GLMM construction, frequentist or
Bayesian, with unspecified reference distribution. The model includes popular
parametric models such as the Poisson or negative-binomial GLMM as limiting cases.

Repeated count outcomes in biomedical research arise when recurrent
events are recorded over time. Investigators may wish to characterize how event rates differ by treatment
and other covariates, how their temporal patterns evolve over time during follow-up and
differ across experimental conditions, and how much subjects differ after
accounting for covariates and time. These questions require a
structured and interpretable model for the conditional mean. But recurrent
count data frequently depart from standard parametric distributional
assumptions. Poisson regression imposes equality of the conditional mean and
variance, while negative-binomial models accommodate overdispersion through a
particular parametric mean--variance relationship. Generalized linear mixed
models (GLMMs) introduce subject-specific random effects to account for
repeated observations, but retain a parametric response family
conditional on the covariates and random effects. Thus, even when the
longitudinal mean structure is scientifically appropriate, inference and
prediction can remain sensitive to assumptions about zeros, dispersion,
skewness, or tail behavior.

Our motivating application concerns recurrent seizure counts measured on
rats over two recording periods after induced status epilepticus. The data exhibit strong overdispersion,
a substantial number of days with zero counts, high between-rat
heterogeneity, and visibly different temporal patterns across periods. The
scientific questions concern how seizure burden evolves within each period,
how much higher it is in the chronic period and how this increase varies over
time, whether the temporal pattern differs between periods beyond a shift in
level, how much rats differ, and how frequent seizure-free and high-burden days are.

To address these challenges, we develop \textit{fc}DPGLM as a semiparametric Bayesian
GLMM for functional count data that builds on the
Dirichlet process generalized linear model (DPGLM) of \citet{alam2026dpglm}.
The \textit{fc}DPGLM embeds the semiparametric DPGLM in a
functional mixed-effects model with smooth condition-specific trajectories,
exposure adjustment, and subject-specific random effects. The mean is thus
specified through an interpretable GLM-style regression. The remaining details of the conditional count distribution are learned from data.

The \textit{fc}DPGLM builds on a line of semiparametric regression models that
retain the GLM mean structure while leaving the outcome distribution
unspecified. \citet{rathouz2009generalized} introduced the semiparametric
GLM (SPGLM), with a nonparametric reference distribution replacing the
parametric family in the usual GLM; they developed maximum likelihood for
outcomes with finite support. \citet{huang2012proportional} cast the model as a mean-parameterized
proportional likelihood ratio model and estimated it by empirical likelihood,
\citet{huang2014joint} established consistency and asymptotic normality of the
joint estimator of the regression parameters and the reference distribution,
and \citet{wurm2018semiparametric}, who named the model the generalized
linear density ratio model, developed a scalable maximum-likelihood algorithm
and the \texttt{gldrm} package. The model has also been used for
outcome-dependent sampling designs \citep{maronge2023generalized}. Several
Bayesian variations of semiparametric GLMs were developed. The Dir-GLM \citep{alam2025dirglm} placed a Dirichlet
distribution prior on the reference distribution for outcomes with finite
support, such as scored ordinal responses, yielding posterior uncertainty for
the reference distribution and for derived functionals such as exceedance
probabilities. The DPGLM \citep{alam2026dpglm} replaced it with a normalized completely
random measure
\citep{kingman1967completely,regazzini2003distributional,james2009posterior,lijoi2010models}.

Other methodological developments have addressed different components of
this problem. Zero-inflated \citep{hall2000zero} and
Conway--Maxwell--Poisson \citep{sellers2010flexible} models and their
mixed-model versions enlarge the parametric family for counts. Bayesian nonparametric approaches have relaxed the
random-effects distribution while keeping a parametric response
\citep{kleinman1998semiparametric}, or modeled conditional
distributions through mixtures
\citep{canale2011bayesian}, in which case the regression
mean is no longer a model parameter and mean contrasts must be recovered
indirectly. Generalized estimating equations
\citep{liang1986longitudinal} and copula-based regression \citep{song2009joint},
including a copula DPGLM \citep{yuan2026copula},
are marginal models: the former gives mean inference robust to the response
distribution but no distributional summaries or predictions, and the latter
represents within-subject dependence explicitly. Recurrent-event
studies such as ours instead call for a conditional model, as in a GLMM,
because they ask how the event rate and its distribution behave for a given
subject and how much subjects differ from one another. Functional-data and recurrent-event methods
\citep{morris2015functional,cook2007statistical} model
mean or intensity trajectories but typically rely on Poisson or
mixed-Poisson assumptions when distributional features are of interest.
Data analysis for the motivating rat epilepsy study requires several features
beyond these models, including a mean regression with interpretable parameters,
conditional inference for subjects through random effects, and a conditional
count distribution without restriction to a parametric family.

\section{A Study of Epileptic Seizures in Rats}
\label{sec:application}

We consider data analysis for an animal study
of epileptic seizures in rats, conducted at the University of Colorado with a chemoconvulsant (pilocarpine) model of epilepsy \citep{qiu2026predicting}. Status epilepticus was induced, and spontaneous seizures were then recorded in
$R=21$ rats in two periods: an early period, during
epileptogenesis, beginning three days after status epilepticus (\emph{Stage~1}), and
a chronic period beginning eight weeks after status epilepticus (\emph{Stage~2}).
Seizures were not recorded between the two periods.
By design, seizure activity is expected to be higher and more stable across
days in the chronic period, once epilepsy is established. 
Seizures were identified by offline analysis of continuous video-EEG (24 hours
per day) as high-amplitude (at least three times baseline), evolving epileptiform
discharges of at least 5~Hz lasting at least 10~seconds, with video used to exclude
movement artifacts \citep{lam2017calpain,lam2019calpain}.

The original data contain the start time, duration, and type of each
recorded seizure, together with tracking intervals describing when each rat
was under observation. We aggregate the event and tracking records to the rat-day level, counting all seizures regardless of type (Appendix~\ref{sec:supp-data}). For
observation $i$, let $Y_i$ denote the total number of seizures, let $e_i$
denote the corresponding observed tracking time expressed in days, let
$d_i\in\{1,2\}$ denote the stage, and let $r(i)$ identify
the rat. The resulting analysis contains $n=618$ rat-day observations.

Time within each stage, $t_i$, is measured in days, defined as
consecutive 24-hour intervals from the start of each rat's recording in that
stage. The periods were designed to last
$12$ and $16$ days \citep{qiu2026predicting}, but recording continued longer
for some rats; we use all observed days, up to $13$ in Stage~1 and $21$ in Stage~2, so the two
stages overlap on days $1$--$13$, which we call the common
follow-up period. Because rats are not always observed for a full 24 hours,
the observed time $e_i$ enters the model as an exposure.

The daily seizure counts exhibit substantial heterogeneity and are far from a
simple Poisson model. Overall, the mean count is $10.20$ but the variance is $174.74$;
$29.3\%$ of rat-days have no seizures, and the largest count is 78. The stages differ
markedly: mean daily counts are $4.49$ and $14.32$, and zero-count proportions
$45.6\%$ and $17.5\%$, in Stages~1 and~2. We call a day high-burden if it has at
least $20$ seizures, about twice the overall mean and the 80th percentile of
daily counts; $21\%$ of rat-days are high-burden ($5\%$ in Stage~1 and $31\%$ in
Stage~2). Part of the excess variance reflects differences between stages, days
and rats; whether overdispersion, excess zeros and a long upper tail remain
even after accounting for them is a question about the count distribution, not the
mean. This motivates a model that specifies the mean but learns the remaining
count distribution from the data.

\begin{figure}[!ht]
    \centering
    \includegraphics[width=\linewidth]{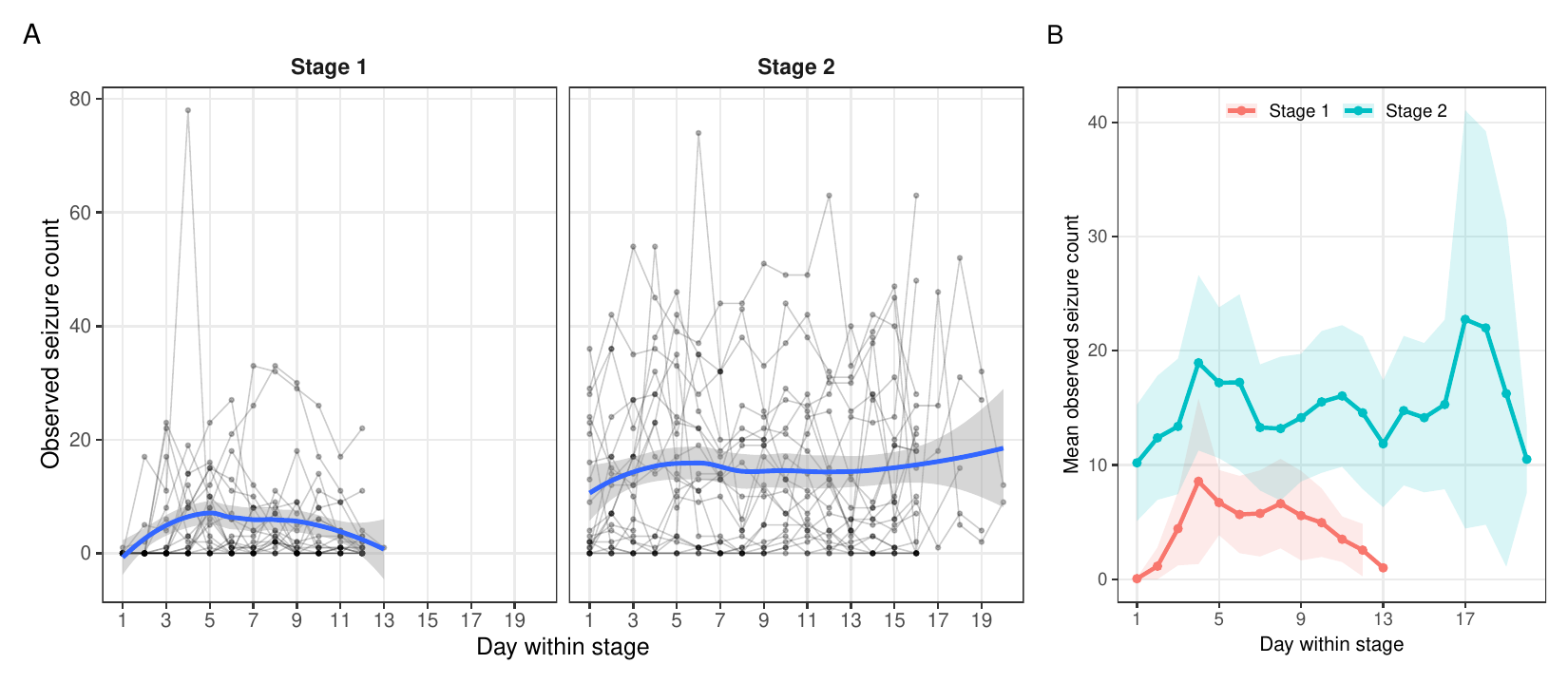}
    \caption{Exploratory summaries of the recurrent seizure data.
(A) Observed daily seizure counts by day since the start of each stage. Thin lines connect observations from the same rat, and solid curves are
stage-specific LOESS smooths. (B) Observed mean daily seizure counts by day for the two stages. Points
 denote empirical daily means and shaded regions show mean $\pm 1.96$ standard errors.
The wider bands at later Stage~2 days reflect
fewer observations and substantial between-rat variability.}
    \label{fig:rat-eda}
\end{figure}

Figure~\ref{fig:rat-eda} shows some exploratory summaries of the longitudinal
count trajectories. Seizure frequency rises
and then falls in Stage~1, whereas Stage~2 starts substantially higher and
stays elevated and stable over the common follow-up period. Individual
trajectories show considerable between-rat heterogeneity (Figure~\ref{fig:supp-rat-trajectories}). A single stage effect therefore cannot describe how the difference between
stages changes over time.

Our analysis is organized around five primary scientific questions.
First, how does the expected daily seizure count evolve over time within
each stage? Second, by how much does the
seizure rate increase from the early to the chronic period, and how does the
increase vary over the days of follow-up? For a typical rat (random effect
$a=0$) observed for a full day, the principal absolute contrast is
\begin{equation}
  \Delta(t)
  =
  \E(Y\mid d=2,t,e=1,a=0)
  -
  \E(Y\mid d=1,t,e=1,a=0).
  \label{eq:target-delta}
\end{equation}
We also consider the corresponding rate ratio and averages of these
contrasts over the common follow-up period. Third, does the
temporal pattern of seizure activity change between periods, beyond a shift
in level, as expected if seizure activity stabilizes in the chronic period? 
Fourth, how much persistent between-rat heterogeneity remains after
accounting for stage and time? Fifth, how often are
rat-days seizure-free, and how likely is a high-burden day ($\geq 20$
seizures) as a function of the expected rate?
The first four questions concern the conditional mean, whereas the fifth
depends on the count distribution beyond its mean.

\section{The \textit{fc}DPGLM Model}
\label{sec:model}

The proposed \textit{fc}DPGLM model separates the specification of the
conditional mean from that of the remaining distributional details: the mean
trajectory is modeled by a GLM regression structure, and the conditional count
distribution by exponential tilting of an unknown reference measure, for which
we introduce a nonparametric Bayesian prior.

\subsection{Semiparametric mean regression by exponential tilting}
\label{sec:tilting}
We build on the DPGLM
\citep{alam2026dpglm}, which we adapt here to count responses. 
Let $y_i$ denote a count response with covariates $x_i$, and let
$\mu_i=\mu(x_i)$ denote the desired conditional mean
imposed by the GLM structure including regression on covariates and possible random effects as in standard GLMs.  We will introduce more details and a specific example later with the discussion of recurrent counts (Section~\ref{sec:functional-model}).
Consider a nonnegative measure $\lambda$ on the sample space 
$\mathcal{Y}=\mathbb{N}_0=\{0,1,\ldots\}$ for the counts.
We define a sampling model for $y_i$ as
\begin{equation}
  p_i(y\mid\theta_i,\lambda) \propto
  \exp(\theta_i y)\lambda(dy)
  \label{eq:tilted-model}
\end{equation}
with log normalization constant
$ %
  b(\theta;\lambda)
  =
  \log\left\{
    \int_{\mathcal{Y}}\exp(\theta y)\lambda(dy)
  \right\}.
$ %
The first two derivatives with respect to $\th$ determine the moments
$b'(\theta;\lambda)=E_{\theta,\lambda}(Y)$ and
$b''(\theta;\lambda)=\Var_{\theta,\lambda}(Y)$.
Under the GLM structure we define 
the conditional mean $\mu_i$ directly, as mentioned above, implying $\theta_i$ through
\begin{equation}
  b'(\theta_i;\lambda)=\mu_i=g^{-1}(x_i^\top\beta),
  \label{eq:mean-constraint}
\end{equation}
for a given link function $g$, covariates $x_i$ and regression coefficients $\beta$.
Provided the reference measure is not concentrated at a single point,
$b''(\theta;\lambda)>0$, so $b'(\theta;\lambda)$ is strictly increasing and
the solution of \eqref{eq:mean-constraint}, when it exists, is unique and determines $\theta_i$ as a derived parameter.
The regression model determines $\mu(x_i)$ as the feature of primary scientific
interest, whereas $\lambda$ determines distributional features that are not
fixed by the mean. This is the construction of the semiparametric GLM of
\citet{rathouz2009generalized}, also known as the density-ratio representation of the response distribution \citep{wurm2018semiparametric}. In contrast to a Poisson or
negative-binomial GLM, however, specifying $\mu_i$ does not simultaneously
impose a particular variance function or a fixed relationship between the
mean and other features of the count distribution. Thus, overdispersion,
excess or reduced mass at zero, skewness, and tail behavior can be learned from the data.

\subsection{A gamma-CRM model for the count distribution}
\label{sec:crm}

To complete the Bayesian inference model, we place a gamma completely random measure
(CRM) \citep{kingman1967completely} prior on $\lambda(\cdot)$. Because the outcome space is discrete, the CRM
can be represented directly by independent random masses over $\mathcal{Y}\subseteq\mathbb{N}_0$,
\begin{equation}
  \lambda
  =
  \sum_{k=0}^{\infty} w_k\delta_k,
  \qquad
  w_k \stackrel{\mathrm{ind}}{\sim}
  \GammaDist\{\alpha G_0(k),1\},
  \label{eq:crm}
\end{equation}
where $G_0$ is a probability mass function on $\mathbb{N}_0$ and
$\alpha>0$ is a concentration parameter. Normalizing $\lambda$ implies a random probability measure,
$f_0(k)=w_k/\sum_{\ell=0}^{\infty}w_\ell$, $k\in\mathbb{N}_0$. In general, a
normalized CRM is known as a normalized random measure (NRM)
\citep{regazzini2003distributional}. Normalizing a gamma CRM gives the Dirichlet
process (DP) \citep{ferguson1973bayesian,lijoi2026ferguson}, $f_0\sim\mathrm{DP}(\alpha,G_0)$,
with base distribution $G_0$ and concentration $\alpha$. Under \eqref{eq:tilted-model} and \eqref{eq:crm}, the conditional probability mass function is
\begin{equation}
  p_i(k) \equiv p(y_i=k\mid \th_i,w)
  \propto {w_k\exp(\theta_i k)},
  \qquad k\in\mathbb{N}_0,
  \label{eq:count-likelihood}
\end{equation}
where $\theta_i=\theta(\mu_i,w)$ is determined implicitly by
$\mu_i = \sum k\, p_i(k)$. In the application
$G_0$ is negative binomial.

Two properties of this construction guide its use. For computation the
count space is truncated to $\mathbb N_{0,K}=\{0,\ldots,K\}$
(Section~\ref{sec:computation}). For a
probability mass function $\pi$ on $\mathbb N_{0,K}$ with full support,
$\pi(k)>0$, and for $0<\mu<K$, let $Q^\pi_\mu$ denote the
exponential tilt of $\pi$ with mean $\mu$, $Q^\pi_\mu(k)\propto\pi(k)e^{\theta k}$
with $\theta$ chosen so that $\sum_kkQ^\pi_\mu(k)=\mu$. We apply this tilt both for the reference distribution $f_0$ and for the base distribution
$G_0$. In particular, $p_i=Q^{f_0}_{\mu_i}$. The first property identifies the
parametric model at the center of the prior, for any choice of $G_0$.

\begin{proposition}[Parametric limit]\label{prop:limit}
Let $G_0(k)>0$ for all $k\in\mathbb N_{0,K}$. For fixed $\mu_i$, as $\alpha\to\infty$, $p_i\to Q^{G_0}_{\mu_i}$
in probability for every $i$. The \textit{fc}DPGLM likelihood converges to
that of the mixed model with conditional distributions $Q^{G_0}_{\mu_i}$. If
$G_0$ is a power-series distribution, $G_0(k)\propto g(k)q^k$, then
$Q^{G_0}_\mu$ is the member of the same family with mean $\mu$, truncated to
$\mathbb N_{0,K}$. In particular, a Poisson base distribution yields the
Poisson GLMM, and a negative-binomial base distribution with size $r$
yields the negative-binomial GLMM with size $r$.
\end{proposition}

\noindent Thus the base distribution chooses the parametric model toward
which the conditional distributions are shrunk, and $\alpha$ controls the
strength of that shrinkage, with finite $\alpha$ allowing departures from it. The next result states identifiability constraints, that is, which features of the model the data determine.

\begin{proposition}[Identification]\label{prop:ident}
Consider $\lambda=\sum_kw_k\delta_k$ and $p_i(k)\propto w_k\exp(\theta_ik)$ as in
\eqref{eq:count-likelihood}. For any $c>0$ and $t\in\mathbb R$, the weights
$w'_k=c\,e^{tk}w_k$ and tilting parameters $\theta_i'=\theta_i-t$ imply the same
likelihood $p_i$. That is, the family $\{Q^{f_0}_\mu:0<\mu<K\}$ of
conditional distributions indexed by the mean is invariant under these
transformations.
\end{proposition}

\noindent The reference measure is therefore identified only up to scale
and exponential tilt, but the quantities of scientific interest are
identified. The mean functionals depend only on the regression parameters
and random effects. Distributional summaries, such as the variance function
$V(\mu;f_0)=\sum_k(k-\mu)^2Q^{f_0}_\mu(k)$, the dispersion index $V(\mu;f_0)/\mu$,
the zero probability $\Pr(Y=0\mid \mu)=Q^{f_0}_\mu(0)$ and the exceedance
probability $\Pr(Y\geq c\mid\mu)=\sum_{k\geq c}Q^{f_0}_\mu(k)$, are functionals
of the invariant family; so are their values at a particular mean, such as
the typical-rat curve $m_d(t)$ or the fitted means $\mu_i$, and the posterior
predictive distribution. However, the prior is not invariant, so along the
unidentified directions the posterior is governed by the prior
(compare also Proposition~\ref{prop:mass}, in Section~\ref{sec:computation}). The
concentration parameter $\alpha$, which we also report, describes the
prior rather than an identified feature of the data distribution, and its
posterior should therefore be read as the strength of shrinkage toward $G_0$. Proofs of Propositions~\ref{prop:limit} and~\ref{prop:ident} and the derivation of
Remark~\ref{rem:score} below are given in Appendix~\ref{sec:supp-proofs}.

\subsection{Functional regression for recurrent counts}
\label{sec:functional-model}

We next embed the semiparametric count distribution in a functional
mixed-effects model appropriate for the recurrent seizure data, with $Y_i$,
$e_i$, $d_i$, $t_i$ and $r(i)$ as defined in Section~\ref{sec:application}.
We use a GLMM framework with rat-specific random effects, and a flexible
spline regression to represent longitudinal profiles.
We model the conditional mean as
\begin{equation}
\begin{split}
  \mu_i
  &=
  E\{Y_i\mid d_i,t_i,e_i,a_{r(i)}\} \\
  &=
  e_i\exp\big[
      \beta_0
      +\beta_{\mathrm{stage}}1(d_i=2)
      +B(t_i)^\top\gamma_1 1(d_i=1)
      +B(t_i)^\top\gamma_2 1(d_i=2)
      +a_{r(i)}
  \big],
\end{split}
\label{eq:mean-model}
\end{equation}
where $B(t)$ is a natural cubic spline basis. Stage-specific spline
coefficients $\gamma_1$ and $\gamma_2$ allow the temporal trajectories to
differ flexibly between the two stages. The exposure $e_i$
enters as an offset, so the regression describes a daily event rate while
retaining partially observed rat-days. 

The inference model is completed with a
random effects distribution and priors. Rat-specific random intercepts are $a_r\sim N(0,\sigma_a^2), \ r=1,\ldots,R$, which account for persistent between-rat heterogeneity. Conditional on the
random effects, repeated observations are assumed independent.
Under the model, the conditional count distribution of $Y_i$
depends on the covariates and the random effect only through $\mu_i$: it is
the tilt $Q^{f_0}_{\mu_i}$ of the reference distribution $f_0$ (which itself is not identified and is not reported
--- see Proposition~\ref{prop:ident}). We place a
half-Cauchy$(0,1)$ prior on $\sigma_a$ \citep{gelman2006prior}, so that
uncertainty in the degree of between-rat heterogeneity is propagated into
all posterior summaries.

For the regression coefficients, we use weakly informative priors for the
intercept and stage effect, $\beta_0,\beta_{\mathrm{stage}}\sim N(0,10^2)$, and a
hierarchical prior for the spline coefficients. Writing each stage-specific
spline term in a basis that is orthonormal over the observed rat-days of that
stage, we take
  $\gamma_d\given\tau_d\sim N(0,\tau_d^2I)$ and $\tau_d\sim\text{half-Cauchy}(0,1), \ d=1,2$,
so that the amount of shrinkage of each stage's temporal trajectory is
learned from the data. Finally, the concentration parameter of the gamma CRM is given the prior $\alpha\sim\GammaDist(1,0.1)$ (shape and
rate), %
so that the data determine the strength of shrinkage toward the parametric model
generated by $G_0$ (Proposition~\ref{prop:limit}).

\subsection{Inference}
\label{sec:inference}

Given the reference measure, inference for the regression has a familiar
form. Let $\eta_i=\log(\mu_i/e_i)$ denote the linear predictor in
\eqref{eq:mean-model} and $V_i=b''(\theta_i;\lambda)$ the variance of $p_i$.

\begin{remark}[Score and information]\label{rem:score}
Holding the CRM weights fixed,
\[
  \frac{\partial\log p_i(y_i)}{\partial\eta_i}=\frac{(y_i-\mu_i)\,\mu_i}{V_i},
  \qquad
  E\left\{\left(\frac{\partial\log p_i(Y_i)}{\partial\eta_i}\right)^2\right\}
  =\frac{\mu_i^2}{V_i},
\]
where the expectation is under $p_i$.
\end{remark}

\noindent Given the reference measure, the score equations for the
regression coefficients, $\sum_ix_i(y_i-\mu_i)\mu_i/V_i=0$, are therefore the
quasi-likelihood equations of a log-link GLM with variance function
$V(\mu;f_0)$, which is learned rather than assumed. They have mean zero
whenever the mean model is correct, whatever the reference measure, so an inaccurate reference measure affects
the efficiency of mean inference rather than the centering of its estimating
equations. This argument is conditional on the random effects and the
reference measure; the simulations of Section~\ref{sec:simulation} confirm
the resulting robustness of the mean functionals for the full posterior. The
same quantities give the gradient and preconditioner of the sampler in
Section~\ref{sec:computation}.

Of particular scientific interest is the stage-specific mean trajectory
for a typical rat ($a=0$) observed for a full day ($e=1$),
\begin{equation}
  m_d(t)
  =
  E\{Y\mid d,t,e=1,a=0\},
  \qquad d=1,2.
  \label{eq:stage-curve}
\end{equation}
In this notation, the time-varying count difference \eqref{eq:target-delta} and rate ratio between stages are
\begin{equation}
  \Delta(t)=m_2(t)-m_1(t), \quad RR(t)=\frac{m_2(t)}{m_1(t)}.
  \label{eq:delta-rr}
\end{equation}
To separate differences in the temporal shape of the trajectories from
differences in their overall level, we also use the centered shape contrast
\begin{equation}
  S(t)=\Bigl\{\log m_2(t)-\overline{\log m_2}\Bigr\}
       -\Bigl\{\log m_1(t)-\overline{\log m_1}\Bigr\},
  \label{eq:shape}
\end{equation}
where $\overline{\log m_d}$ is the average of $\log m_d(t)$ over the common
follow-up period. The contrast satisfies $S(t)\equiv0$ exactly when the two
log-mean trajectories differ by a constant, that is, when the stages differ
in level but not in shape. We assess $S(t)$, and the other curves, with
pointwise and simultaneous (sup-$t$) %
$95\%$ credible bands (Appendix~\ref{sec:supp-bands}).

\subsection{Posterior simulation}
\label{sec:computation}

We implement posterior inference by way of posterior Markov chain Monte Carlo (MCMC) simulation. We truncate the count space
$\mathbb{N}_0$ at an upper bound $K$ well beyond the largest observed count and work on
$\mathbb N_{0,K}=\{0,\ldots,K\}$, truncating and
renormalizing the base distribution $G_0$ accordingly. 
Because the mean constraint has a solution on $\mathbb N_{0,K}$ only for
$0<\mu_i<K$, parameters implying $\mu_i$ outside this range have zero prior
probability; in the seizure analysis this restriction was never violated by any
proposals generated in Metropolis--Hastings style transition probabilities (largest
fitted mean $60$, against $K=108$).

MCMC posterior simulation alternates between updates of the regression and random-effect
parameters and updates of the reference measure. Given $\lambda$, updating the
regression parameters proceeds as in usual GLMs; compare
Remark~\ref{rem:score}. Updating the CRM requires some more discussion. For a given
set of CRM weights and regression parameters, each $\theta_i$ is obtained
by solving the mean constraint on $\mathbb N_{0,K}$. 

For the CRM update, using the identity
$1/T_i(w)=\int_0^\infty \exp\{-u_iT_i(w)\}\,du_i$ with
$T_i(w)=\sum_{k \in \mathbb N_{0,K}} w_k\exp(\theta_i k)$,
we introduce auxiliary variables
\begin{equation}
  u_i\mid w,\theta_i
  \sim
  \GammaDist\{1,T_i(w)\},
  \label{eq:auxiliary-u}
\end{equation}
where the second argument denotes the rate. 
Defining
$\psi(k)=\sum_{i=1}^n u_i\exp(\theta_i k)$, the
augmentation restores conditional gamma conjugacy for the CRM
weights when the tilting parameters are held fixed. In particular, letting $n_k=\sum_i 1(Y_i=k)$ and fixing $\theta_i$,
\begin{equation}
  w_k\mid Y,u,\theta
  \sim
  \GammaDist\left[
    \alpha G_0(k)+n_k,\,
    1+\psi(k)
  \right],
  \qquad k \in \mathbb N_{0,K}.
  \label{eq:crm-proposal}
\end{equation}
If the $\theta_i$ were fixed, \eqref{eq:crm-proposal} would be the exact
full conditional update for $w_k$. Instead we use \eqref{eq:crm-proposal} to generate
proposals $w_k^\star$ for a Metropolis--Hastings transition probability, substituting currently imputed
$\theta_i$ in \eqref{eq:crm-proposal}. In the proposed model, however,
$\theta_i$ is itself a deterministic function of the conditional mean $\mu_i$ and
the CRM weights. We therefore recompute the
$\theta_i$ under the proposed CRM, and apply a Metropolis--Hastings
correction to target the exact posterior under the computational truncation (Appendices~\ref{sec:supp-augmentation} and~\ref{sec:supp-crm}).

In practice we update the CRM weights in several random blocks of the
support, each by the conjugate proposal \eqref{eq:crm-proposal} restricted to
the block with the corresponding exact Metropolis--Hastings correction;
updating a block changes the tilting parameters less than updating all
weights at once, which raises the acceptance rate. Because only relative
weights enter the likelihood, the total mass of the CRM is not informed by
the data, and a sampler that changes it only through the block updates
mixes slowly. Proposition~\ref{prop:mass} gives an exact update along this
direction.
\begin{proposition}\label{prop:mass}
Let $M=\sum_{k\in\mathbb N_{0,K}}w_k$, so that $f_0=w/M$, and let $v=Mu$. Under the
truncated prior \eqref{eq:crm} and the augmentation \eqref{eq:auxiliary-u},
(i) each $\theta_i$ depends on $w$ only through $f_0$, and (ii) conditionally
on $(f_0,v,\beta,a,\alpha)$, where $a_r$ are the random effects in
\eqref{eq:mean-model}, and the data, $M\sim\GammaDist(\alpha,1)$. Hence
drawing $M^\star\sim\GammaDist(\alpha,1)$ and setting
$w\leftarrow(M^\star/M)\,w$ and $u\leftarrow(M/M^\star)\,u$ is an exact Gibbs
step for the augmented posterior.
\end{proposition}
The proof is %
given in Appendix~\ref{sec:supp-mass}. Part (ii) also shows that the data
inform $\alpha$ only through the normalized weights $f_0$, which is why a
joint Metropolis--Hastings move of $\alpha$ and the weights mixes poorly.
Given the weights, $\alpha$ is updated by slice sampling from its full
conditional \citep{neal2003slice}. The regression
coefficients and rat-specific random effects are updated jointly by a
Metropolis-adjusted Langevin step preconditioned by the Fisher information
\citep{girolami2011riemann}, whose gradient and Fisher information are given
by Remark~\ref{rem:score}, and $\sigma_a$ and
the $\tau_d$ are updated by conjugate draws using the inverse-gamma mixture
representation of the half-Cauchy distribution. Details of all updates and
their acceptance ratios are provided in Appendix~\ref{sec:supp-mcmc}.

\section{Simulation Studies}
\label{sec:simulation}

The simulations address three questions. First, can \fcdpglm\ estimate the functional mean estimands of
\eqref{eq:stage-curve}--\eqref{eq:shape} as accurately as parametric count mixed
models, both when a parametric competitor is correctly specified and when it
is not?  Second, does \fcdpglm\ recover the distributional features that are not
determined by the mean, and does it predict new counts well? Third, is the
simultaneous-band assessment of a difference in temporal shape (defined in Section~\ref{sec:inference} and applied in Section~\ref{sec:realdata}) correctly calibrated? We follow the ADEMP
framework of \citet{morris2019using}, which specifies the aims,
data-generating mechanisms, estimands, methods and performance measures of a
simulation study.

\subsection{Data-generating mechanisms}
\label{sec:sim-design}

The study has a $2\times2\times2$ factorial design: two conditional count
distributions, two sets of mean curves and two sample sizes (number of rats), giving
eight settings with $200$ replicates per setting (Table~\ref{tab:sim-design}). 

\begin{table}[!t]
  \centering
\caption{Simulation design: a $2\times2\times2$ factorial with $200$
replicates per setting, each data set analyzed by \fcdpglm\ and by Poisson
and negative-binomial mixed models. The shape-difference settings give the
mean, distributional and prediction results (Table~\ref{tab:sim-results}) and the power of the shape test; the null-shape settings
give its false-positive rate. For more details, see Section~\ref{sec:sim-design} and Appendix~\ref{sec:supp-sim-design}.}
  \label{tab:sim-design}
  \small
  \renewcommand{\arraystretch}{1.15}
  \begin{tabular}{@{}l l >{\raggedright\arraybackslash}p{5.4cm} >{\raggedright\arraybackslash}p{4.2cm}@{}}
    \toprule
    Factor & Level & Data-generating truth & Role \\
    \midrule
    Count distribution & Negative binomial & Negative binomial with mean
      $\mu_i$ and size $\phi^\star=0.98$ & NB mixed model correct \\
     & Semiparametric & Exponential tilt, to mean $\mu_i$, of the reference distribution estimated from the seizure data & Parametric mixed models misspecified \\
    \addlinespace
    Mean curves & Shape difference & Curves estimated from the seizure data;
      $\max_t|S(t)|=1.8$ & Accuracy; power of the shape test \\
     & Null shape & Stages differ in level only; $S(t)=0$ up to $0.0012$
      & False-positive rate of the shape test \\
    \addlinespace
    Number of rats & $R=21$ & Full design of the seizure study: all $21$ rats, with their observed days and exposures ($618$ rat-days)
      & Size of the seizure study \\
     & $R=10$ & Random subset of $10$ of the $21$ rats, with their observed days and exposures, redrawn in each replicate (about $290$ rat-days)  & Smaller study \\
    \bottomrule
  \end{tabular}
\end{table}
The truth is designed to mimic the seizure data. We fitted the %
SPGLM of \citet{rathouz2009generalized}, with the
mean model \eqref{eq:mean-model}, using the \texttt{gldrm} package
\citep{wurm2018semiparametric}. Because %
\texttt{gldrm} has no random effects, we
entered the rats as fixed effects constrained to sum to zero, so that the
fitted stage-specific mean curves refer to a typical rat ($a=0$). This yields regression coefficients, $21$ rat effects %
and an empirical reference distribution supported on the $54$ distinct
observed counts between $0$ and $78$, which we treat as the true values
$\beta^\star$, $a^\star_r$ and $f_0^\star$ (Appendix~\ref{sec:supp-sim-truth}).
In each replicate the rat effects are drawn with replacement from
$\{a^\star_r\}$. The counts are then generated from one of two conditional distributions:
(i) \underline{Negative-binomial (NB) scenario}: $Y_i$ is negative binomial
  with mean $\mu_i$ and size $\phi^\star=0.98$ (i.e., variance
  $\mu_i+\mu_i^2/\phi^\star$); the
  size $\phi^\star$ was estimated by a negative-binomial GLM with the same mean model.
  Or, (ii) \underline{Semiparametric (SP) scenario}: $Y_i\sim Q^{f_0^\star}_{\mu_i}$, the
  exponential tilt of $f_0^\star$ with mean $\mu_i$ (Section~\ref{sec:crm}),
  so that $\Pr(Y_i=k)\propto f_0^\star(k)e^{\theta_ik}$. This is the
  conditional distribution learned from the seizure data, not a member of a
  parametric family.

\begin{figure}[!ht]
  \centering
  \includegraphics[width=\linewidth]{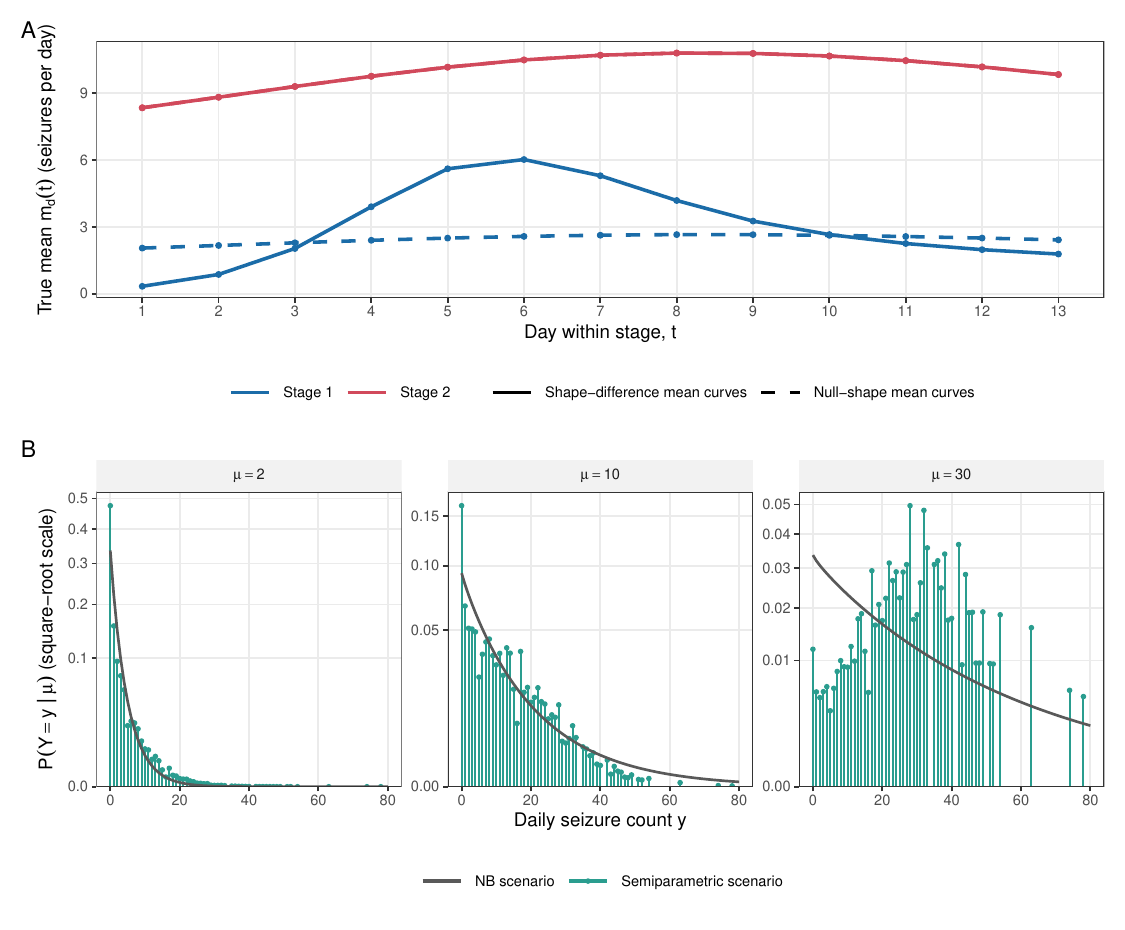}
  \caption{Simulation truth. (A)~Stage-specific mean curves $m_d(t)$ for a typical rat:
shape-difference curves (solid), estimated from the seizure data, and
null-shape curves (dashed), in which the two stages differ in level only;
the Stage~2 curve is the same in both. (B)~Conditional count
distributions $P(Y=y\mid\mu)$ at means $\mu=2$, $10$ and $30$: negative binomial (line) and semiparametric (needles, supported on
the counts observed in the seizure data).}
  \label{fig:supp-sim-truth}
\end{figure}

Because both scenarios share the same regression coefficients and rat
effects, they differ only in the conditional distribution of the counts given
their means. Figure~\ref{fig:supp-sim-truth} shows the true stage-specific mean curves and
the two conditional distributions at three specific means. %

To calibrate the test of a difference in temporal shape, we also generated
data under a \emph{null-shape} truth in which the two stages differ in level
only: the Stage~1 coefficients are replaced so that the true shape contrast is $S(t)=0$ up to a
projection error of at most $0.0012$, leaving the Stage~2 curve and the average log rate ratio
($1.40$) unchanged; the construction is in Appendix~\ref{sec:supp-sim-null} and the curves are in Figure~\ref{fig:supp-sim-truth}(A) and Table~\ref{tab:supp-sim-null-curves}.

In each replicate, a second count is also generated for every rat-day,
independently and from the same conditional distribution, and held out for
assessing prediction.

\subsection{Estimands and performance measures}
\label{sec:sim-estimands}

The estimands mirror the analysis of the recurrent seizure rat study. For a typical rat with one day
of exposure, on the common day grid $t=1,\ldots,13$, they are the
stage-specific mean curves $m_d(t)$ of \eqref{eq:stage-curve}, the difference
$\Delta(t)$ of \eqref{eq:delta-rr} and the shape contrast $S(t)$ of
\eqref{eq:shape}, together with two averages over the grid:
$\bar\Delta=\tfrac{1}{13}\sum_t\Delta(t)$ and the average log rate ratio
$\tfrac{1}{13}\sum_t\log RR(t)$, with $RR(t)$ as defined in \eqref{eq:delta-rr}. %

Distributional estimands not determined by the mean are the dispersion index $V(\mu)/\mu$ and the exceedance probability $\Pr(Y\geq20\mid\mu)$ of a
high-burden day (Section~\ref{sec:application}), both evaluated on a grid of means from
$0.5$ to $30$ (Appendix~\ref{sec:supp-sim-exceed});  because both condition on the mean $\mu$, they
measure the conditional distribution beyond the mean, separately from errors
in the mean model. We also report a typical rat's zero probability  $\Pr(Y=0\mid d,t)$ at each
day $t$ of both stages $d$ (that is, at the mean $m_d(t)$), and the zero probability, conditional variance and $\Pr(Y\geq20)$ averaged over the
rat-days of each data set. Held-out prediction is assessed by the ranked probability score (RPS), a proper scoring rule for counts \citep{gneiting2007strictly}, reported as the excess RPS, the method's RPS minus that of the true distribution, so that zero is perfect.

For each estimand we report bias, root mean squared error (RMSE) and
empirical coverage of nominal $95\%$ intervals. For functions of the day or of
the mean we average over the grid: the integrated RMSE is the square root
of the grid-averaged mean squared error, and coverage is the mean pointwise
coverage. For the curves of Section~\ref{sec:inference} we also report
the coverage of $95\%$ simultaneous (sup-$t$) bands over the common follow-up period.

\subsection{Comparison of inference models}
\label{sec:sim-methods}

\emph{\fcdpglm.} We fit the model of
Sections~\ref{sec:tilting}--\ref{sec:functional-model} using the sampler of Section~\ref{sec:computation}, with four random blocks for the CRM weights. Sensitivity to the prior specification is reported in Appendix~\ref{sec:supp-sim-sens}.
The base distribution $G_0$ is negative binomial with empirical-Bayes mean and size, truncated at $K=\max_iy_i+30$.
Each fit is based on
$3{,}000$ MCMC iterations and discards the first $1{,}000$, during which the step
sizes are adapted. %
We report posterior medians and $95\%$ equal-tailed credible intervals. 
The \fcdpglm\ predictive distribution
averages the tilted reference distribution over $200$ equally spaced posterior
draws.

\emph{Parametric models.} Poisson and NB mixed models were fitted with
\texttt{glmmTMB} \citep{brooks2017glmmtmb} using the same exposure offset,
stage-specific spline basis and random intercept. Point estimates are the
maximum-likelihood estimates of the fixed effects, plugged into the
functionals. For the mean functionals, intervals and simultaneous bands are
obtained from $2{,}000$ draws from the asymptotic normal distribution of the
fixed-effect estimates, with the NB size parameter held at its estimate; for
the distributional functionals we use $1{,}000$ such draws. %
Predictive distributions condition on the predicted rat effects. 

\subsection{Results}
\label{sec:sim-results}

\paragraph{Mean functionals}
Table~\ref{tab:sim-results}, Panel~A, summarizes the mean functionals; results
with ten rats lead to the same comparisons (Table~\ref{tab:supp-sim-mean10}).
When the NB model is correct, \fcdpglm\ is as accurate as the NB mixed model,
with RMSE within $1\%$ for the average log rate ratio and smaller for the other
estimands. Under the semiparametric scenario, \fcdpglm\ is more accurate: the
integrated RMSE for $\Delta(t)$ is $1.96$ against $2.12$, and the RMSE and bias
of the average log rate ratio are $0.102$ and $0.011$ against $0.117$ and
$0.046$. In both scenarios, \fcdpglm\ intervals have close to nominal coverage,
whereas the Poisson mixed model, which ignores overdispersion, gives intervals
far too narrow: coverage $0.54$--$0.56$ for the average log rate ratio and
$0.18$--$0.22$ for the simultaneous band of $S(t)$. Pointwise coverage and the
recovered zero and exceedance probabilities are in Figures~\ref{fig:supp-sim-coverage} and~\ref{fig:supp-sim-distcurves}.

\begin{table}[p]
  \centering
  \caption{Simulation results with $R=21$ rats, shape-difference curves, $200$ replicates per setting. Panel~A, mean functionals: bias, RMSE and coverage (Cov) of $95\%$ intervals for the average log rate ratio; RMSE and Cov for the average difference $\bar\Delta$; integrated RMSE (IRMSE), mean pointwise coverage (Cov) and simultaneous-band coverage (Sim) of the $95\%$ bands for $\Delta(t)$; and IRMSE and Sim for the shape contrast $S(t)$. Panel~B, distributional features and held-out prediction: bias of the zero probability, conditional variance and $\Pr(Y\geq20)$ averaged over the rat-days (true values in parentheses); IRMSE and Cov for $\Pr(Y=0\mid d,t)$ and for $\Pr(Y\geq20\mid\mu)$ (no coverage for the Poisson model, under which $\Pr(Y\geq20\mid\mu)$ is fixed); excess RPS of the held-out counts over the true distribution; bold marks the main contrasts under the semiparametric scenario. Panel~C, dispersion index $V(\mu)/\mu$ of the conditional count distribution: truth (open circles), and mean and $2.5$--$97.5\%$ range of the estimates over the replicates (solid lines and bands), with axis tick labels spaced logarithmically. Results with $R=10$ rats are in Tables~\ref{tab:supp-sim-mean10} and~\ref{tab:supp-sim-dist10} and Figure~\ref{fig:supp-sim-disp10}.}
  \label{tab:sim-results}
  \footnotesize
  \setlength{\tabcolsep}{4pt}
  \resizebox{\linewidth}{!}{%
  \begin{tabular}{llrrrrrrrrrr}
    \toprule
    \multicolumn{12}{l}{\textit{Panel A: Mean functionals}} \\
    \midrule
    & & \multicolumn{3}{c}{Avg.\ log RR} &
      \multicolumn{2}{c}{$\bar\Delta$}
      & \multicolumn{3}{c}{$\Delta(t)$} & \multicolumn{2}{c}{$S(t)$} \\
    \cmidrule(lr){3-5}\cmidrule(lr){6-7}\cmidrule(lr){8-10}\cmidrule(lr){11-12}
    Scenario & Method & Bias & RMSE & Cov & RMSE & Cov & IRMSE &
      Cov & Sim & IRMSE & Sim \\
    \midrule
    NB & \fcdpglm & $-0.012$ & 0.111 & 0.935 & 1.71 &
      0.945 & 1.94 & 0.948 & 0.940 & 0.177 & 0.930 \\
     & NB GLMM & $0.003$ & 0.110 & 0.935 & 1.75 &
      0.940 & 2.05 & 0.935 & 0.925 & 0.184 & 0.945 \\
     & Poisson GLMM & $0.010$ & 0.125 & 0.535 & 1.69 &
      0.900 & 2.07 & 0.829 & 0.660 & 0.216 & 0.180 \\
    \addlinespace
    Semiparametric & \fcdpglm & $0.011$ & 0.102 & 0.945 & 1.80 &
      0.935 & 1.96 & 0.938 & 0.930 & 0.208 & 0.900 \\
     & NB GLMM & $0.046$ & 0.117 & 0.925 & 1.90 &
      0.920 & 2.12 & 0.937 & 0.955 & 0.219 & 0.880 \\
     & Poisson GLMM & $0.039$ & 0.117 & 0.555 & 1.76 &
      0.910 & 1.93 & 0.889 & 0.765 & 0.222 & 0.220 \\
    \bottomrule
  \end{tabular}}

  \vspace{8pt}

  \resizebox{\linewidth}{!}{%
  \begin{tabular}{llrrrrrrrr}
    \toprule
    \multicolumn{10}{l}{\textit{Panel B: Distributional features and held-out prediction}} \\
    \midrule
    & & \multicolumn{3}{c}{Bias, averaged over rat-days}
      & \multicolumn{2}{c}{$\Pr(Y=0\mid d,t)$} & \multicolumn{2}{c}{$\Pr(Y\geq20\mid\mu)$} & \raisebox{-1.1ex}[0pt][0pt]{Excess} \\
    \cmidrule(lr){3-5}\cmidrule(lr){6-7}\cmidrule(lr){8-9}
    Scenario & Method & $\Pr(Y=0)$ & Variance & $\Pr(Y\geq20)$ &
      IRMSE & Cov & IRMSE & Cov & RPS \\
    \midrule
    NB & & (0.225) & (215) & (0.157) & & & & & \\
     & \fcdpglm & $-0.001$ & $9.1$ & $-0.001$ &
      0.045 & 0.95 & 0.014 & 0.96 & 0.129 \\
     & NB GLMM & $-0.001$ & $-8.5$ & $-0.006$ &
      0.044 & 0.92 & 0.004 & 0.96 & 0.124 \\
     & Poisson GLMM & $-0.102$ & $-205$ & $0.017$ &
      0.142 & 0.13 & 0.200 & -- & 1.499 \\
    \addlinespace
    Semiparametric & & (0.298) & (76.6) & (0.218) & & & & & \\
     & \fcdpglm & $-0.012$ & $4.2$ & $-0.008$ &
      0.054 & \textbf{0.92} & 0.013 & \textbf{0.91} & \textbf{0.121} \\
     & NB GLMM & $-0.065$ & $133$ & $-0.065$ &
      0.110 & \textbf{0.38} & 0.113 & \textbf{0.04} & \textbf{0.449} \\
     & Poisson GLMM & $-0.170$ & $-66.4$ & $-0.035$ &
      0.237 & \textbf{0.06} & 0.126 & -- & \textbf{0.853} \\
    \bottomrule
  \end{tabular}}

  \vspace{8pt}

  {\raggedright\textit{Panel C: Dispersion index $V(\mu)/\mu$}\par}
  \vspace{2pt}
  \includegraphics[width=\linewidth]{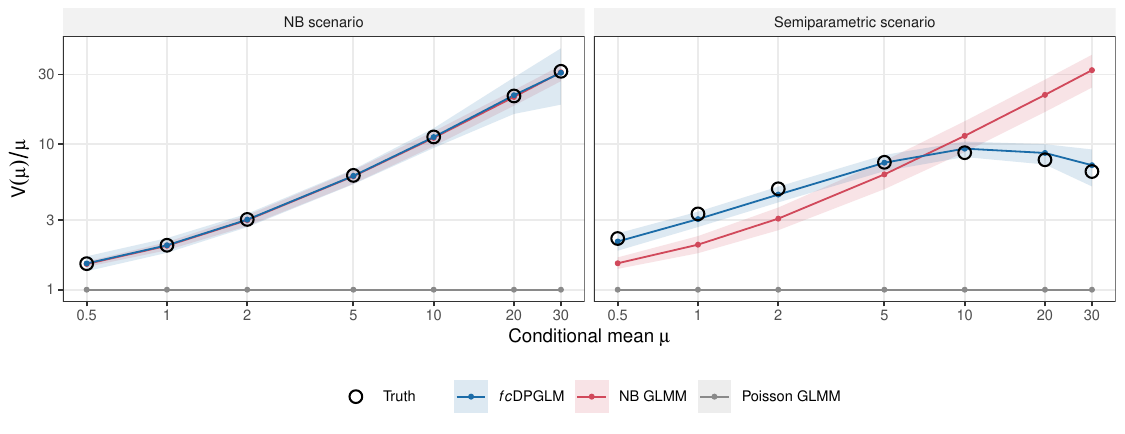}
\end{table}

\paragraph{Testing for a difference in temporal shape}
We declare a shape difference when the $95\%$ simultaneous band for $S(t)$
excludes zero on at least one day, a test of nominal level $5\%$. Under the null-shape truth, \fcdpglm\ rejected the null hypothesis of equal shape in $0.5$--$2.0\%$ of replicates across the four null settings, compared with $5.5$--$7.5\%$ for the NB mixed model and $73$--$80\%$ for the Poisson mixed model. Under the shape-difference curves, the power of \fcdpglm\ was $1.00$ with $21$ rats and, with ten rats, $0.97$ under the NB truth and $0.905$ under the semiparametric truth (NB mixed model: $0.985$--$1.00$). A shape difference detected by \fcdpglm\ is
thus rarely a false positive, at little cost in power.

\paragraph{Distributional features}
Table~\ref{tab:sim-results}, Panels~B and~C, show where the methods differ most. In the semiparametric scenario, the true dispersion index rises with the mean up to about $10$ and then falls, a shape the quadratic variance function of the NB mixed model cannot reproduce. \fcdpglm\ recovers it, with integrated RMSE $0.79$ against $11.3$ for the NB mixed model, which overstates the conditional variance averaged over the rat-days by a factor of about $2.7$. For the typical-rat zero probability $\Pr(Y=0\mid d,t)$, the integrated RMSE of \fcdpglm\ is half that of the NB mixed model, with interval coverage $0.92$ against $0.38$. The upper tail shows the same pattern: for $\Pr(Y\geq20\mid\mu)$, the integrated RMSE is $0.013$
against $0.113$, with coverage $0.91$ against $0.04$. The NB mixed model understates this probability at large means, by $0.13$ at $\mu=20$ and $0.26$
at $\mu=30$ (true values $0.50$ and $0.78$), because its variance function spreads the distribution too widely there. Under the NB scenario, both methods
recover these features about equally well, with coverage $0.92$--$0.96$. 

\paragraph{Prediction}
Under the semiparametric scenario, \fcdpglm\ predicts the held-out counts best
(Table~\ref{tab:sim-results}, Panel~B), with excess RPS $0.121$ against $0.449$
for the NB and $0.853$ for the Poisson mixed model; under the NB scenario,
\fcdpglm\ and the NB mixed model are essentially equal ($0.129$ against
$0.124$). Results with ten rats are in Table~\ref{tab:supp-sim-dist10}.

\section{Recurrent Seizure Study}
\label{sec:realdata}

We fit \textit{fc}DPGLM to the $618$ rat-days with the prior, base distribution and truncation as in the simulation study, using four chains of $30{,}000$ iterations that give $20{,}000$ posterior draws after burn-in and thinning; the potential scale reduction factor $\hat R$ \citep{gelman1992inference} was at most $1.001$ and the effective sample size exceeded $4{,}600$ for every regression coefficient, $\sigma_a$ and $\log\alpha$ (diagnostics in Appendix~\ref{sec:supp-mcmc-diagnostics}). Additional posterior summaries are in Appendix~\ref{sec:supp-additional-results}. 

\subsection{Temporal evolution of seizure burden}
\begin{figure}[!ht]
  \centering
  \includegraphics[width=\linewidth]{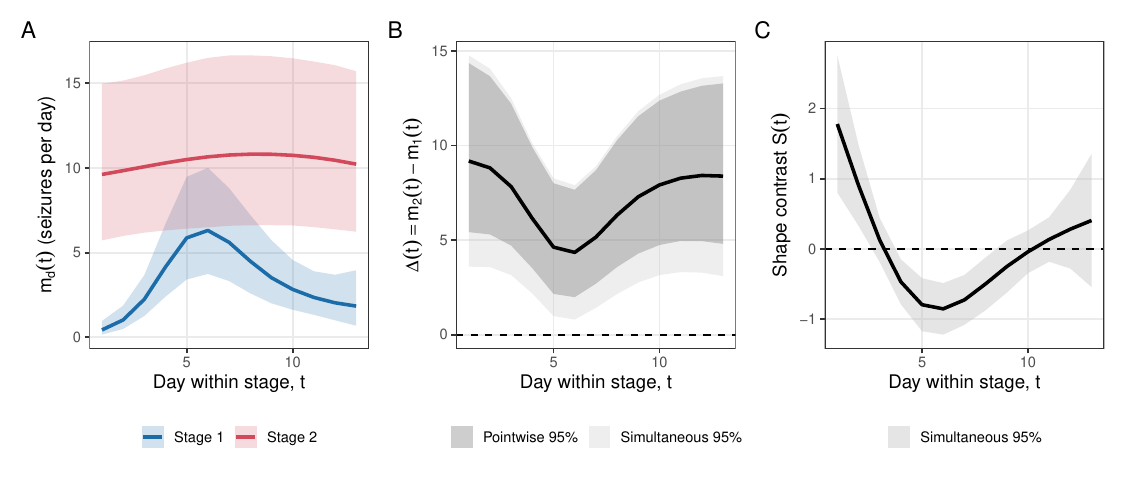}
\caption{Posterior inference for the mean functionals in the seizure study under \textit{fc}DPGLM, for a typical rat ($a=0$) with one day of exposure. (A)~Stage-specific mean trajectories $m_d(t)$: posterior means and pointwise $95\%$ credible intervals. (B)~Posterior mean of the stage difference $\Delta(t)$ with pointwise (dark) and simultaneous (light) $95\%$ credible bands. (C)~Posterior mean of the shape contrast $S(t)$ of \eqref{eq:shape} with a simultaneous $95\%$ credible band; the dashed line at zero corresponds to no difference in temporal shape.}
  \label{fig:case-functionals}
\end{figure}

Figure~\ref{fig:case-functionals}A shows the estimated stage-specific mean
seizure trajectories under \textit{fc}DPGLM for a typical rat ($a=0$) observed
for one full day. The Stage~1 trajectory rises to a peak of $6.31$ $(3.74,10.01)$
seizures per day on day~6 and then declines, whereas the Stage~2 trajectory is
substantially higher, between $9.6$ and $10.8$ seizures per day, and nearly
flat. Values by day, with $95\%$ credible intervals, are in Table~\ref{tab:supp-bydelta}.

\subsection{Time-varying increase in seizure burden}
Figure~\ref{fig:case-functionals}B shows the stage difference $\Delta(t)$ of
\eqref{eq:target-delta}. Its simultaneous $95\%$ band stays above zero
throughout the common follow-up period, so seizure rates are higher in the
chronic period on every day. The increase varies over time and is smallest on
day~6, when Stage~1 activity peaks ($4.35$ $(1.98,7.66)$ seizures per day).
Averaged over the common day grid, the difference is $7.13$ $(4.33,11.01)$
seizures per day, a geometric-mean rate ratio of $4.00$ $(3.25,4.89)$. These
contrasts are for a typical rat; averaging over a new random effect multiplies
the mean trajectories and their difference by $\exp(\sigma_a^2/2)$, estimated as
$1.63$ $(1.30,2.77)$, and leaves rate ratios and shape contrast $S(t)$ unchanged (Appendix~\ref{sec:supp-additional-results}).

\subsection{Differences in temporal shape}
To separate differences in temporal shape from differences in level, we use the
shape contrast $S(t)$ of \eqref{eq:shape}, which centers each stage-specific
log-rate trajectory by its average over the common day grid
(Figure~\ref{fig:case-functionals}C). The simultaneous $95\%$ band excludes zero
on days~1--2, where $S(t)>0$ because Stage~1 starts low relative to its own
average, and on days~4--8, where $S(t)<0$ because Stage~1 peaks. The stages
therefore differ in shape, not only in level, consistent with the more stable
seizure activity expected in the chronic period.

\subsection{Between-rat heterogeneity}
\begin{figure}[!ht]
  \centering
  \includegraphics[width=0.62\linewidth]{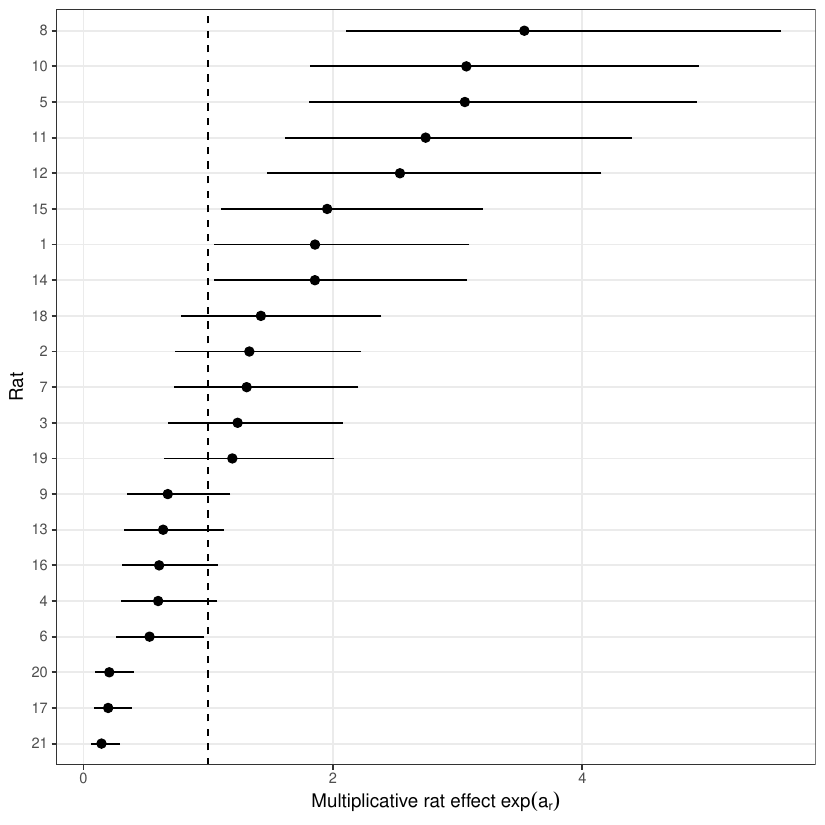}
  \caption{Posterior means and $95\%$ credible intervals of the rat effects
  $\exp(a_r)$ under \fcdpglm{}, ordered by posterior mean. The dashed line at
  one marks no rat effect.}
  \label{fig:supp-rat-effects}
\end{figure}

The random-effect standard deviation has posterior median $\sigma_a=0.99$
$(0.72,1.43)$, and rat effects account for $58\%$ $(47\%,68\%)$ of the variance
of the linear predictor, $\mathrm{Var}(a_r)/\{\mathrm{Var}(a_r)+\mathrm{Var}(x_i^\top\beta)\}$,
computed within each posterior draw over the observed rat-days. The most
affected rat has a seizure rate $30.8$ $(16.9,55.7)$ times that of the least
affected, $\exp(\max_ra_r-\min_ra_r)$. Substantial between-rat heterogeneity
therefore remains after accounting for stage and time. Figure~\ref{fig:supp-rat-effects} shows the rat effects $\exp(a_r)$ relative to
the typical rat.

\subsection{Robustness of the mean conclusions to the count distribution}
\label{sec:compare-parametric}

\begin{figure}[!ht]
  \centering
  \includegraphics[width=\linewidth]{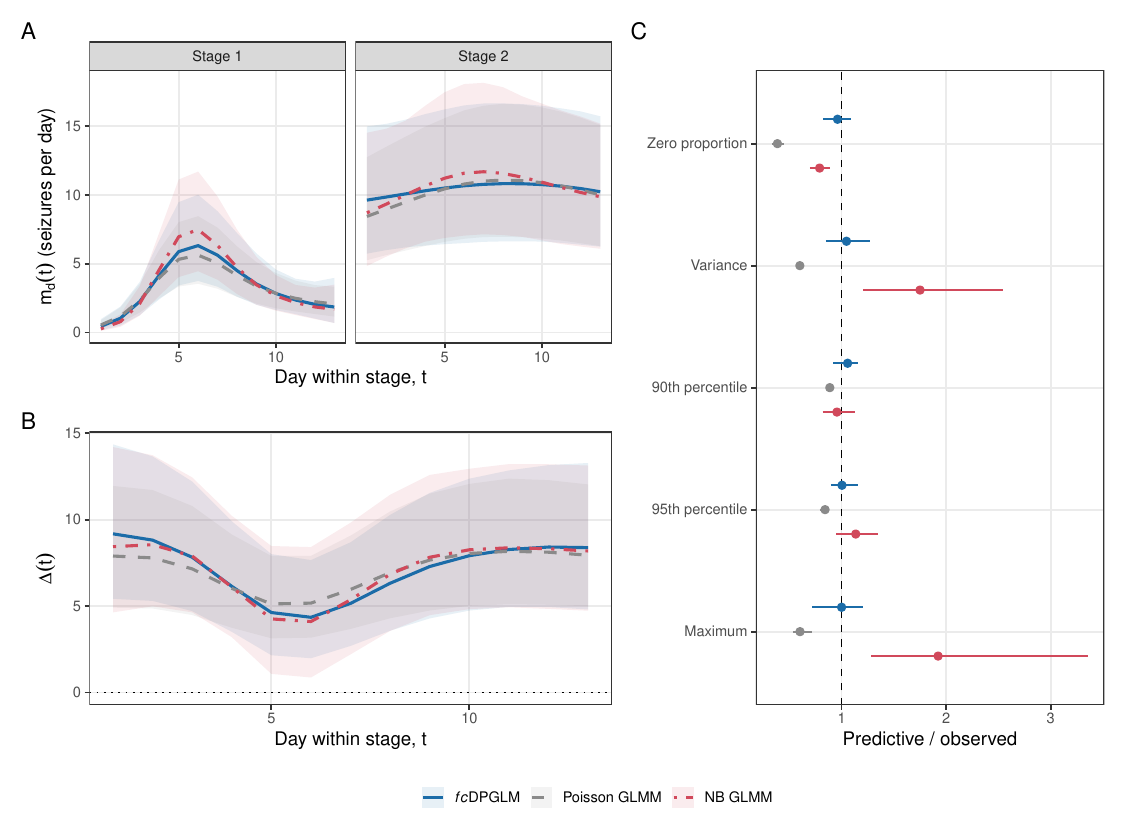}
  \caption{Comparison of \textit{fc}DPGLM with the Poisson and negative-binomial GLMMs in the seizure study. (A)~Stage-specific mean trajectories $m_d(t)$ and (B)~stage difference $\Delta(t)$ for a typical rat with one day of exposure, with pointwise $95\%$ intervals. (C)~Predicted zero proportion, variance, $90$th percentile, $95$th percentile and maximum of the daily counts, with $95\%$ predictive intervals, as ratios to the observed values; the dashed line at one indicates exact agreement.}
  \label{fig:case-comparison}
\end{figure}

To assess whether the conclusions about the conditional mean depend on the
count distribution, we also fit Poisson and negative-binomial GLMMs with the
same exposure offset, spline basis and random intercept as \textit{fc}DPGLM.
The mean trajectories are broadly similar across the three models
(Figure~\ref{fig:case-comparison}A). In Stage~1, the negative-binomial GLMM
gives the most pronounced peak and the Poisson GLMM the least, with
\textit{fc}DPGLM in between; in Stage~2 the three trajectories nearly coincide,
and the intervals overlap substantially throughout.

The stage differences are similarly robust (Figure~\ref{fig:case-comparison}B):
the average difference is about $7.1$ seizures per day under all three models,
and the geometric-mean rate ratio is $4.0$ under \textit{fc}DPGLM, $3.8$ under
the Poisson GLMM and $4.2$ under the negative-binomial GLMM (intervals in Table~\ref{tab:supp-contrasts}). The much narrower Poisson interval reflects its
neglect of overdispersion, not greater information. The conclusions about the
mean are therefore not driven by the count distribution.

\subsection{Seizure-free and high-burden days}
\label{sec:distribution}
The fifth question depends on the count distribution beyond its mean, and here
the models differ markedly. Figure~\ref{fig:case-comparison}C compares each
model's predictive statistics with the observed values; a ratio of one means
exact reproduction. The \textit{fc}DPGLM statistics are all close to the observed values, with the observed value lies inside its
predictive interval for every statistic. The Poisson GLMM falls below the
observed value for every statistic, with too few zeros, too little variability
and too light an upper tail. The negative-binomial GLMM also gives too few
zeros and overstates the variance and the maximum, although its $90$th and
$95$th percentiles are compatible with the observed values, the $95$th tending
high. Predictive summaries are in Tables~\ref{tab:supp-ppc}
and~\ref{tab:supp-ppc-stage} and Figure~\ref{fig:supp-ppc} (Appendix~\ref{sec:supp-ppc}).

\begin{figure}[!ht]
  \centering
  \includegraphics[width=\linewidth]{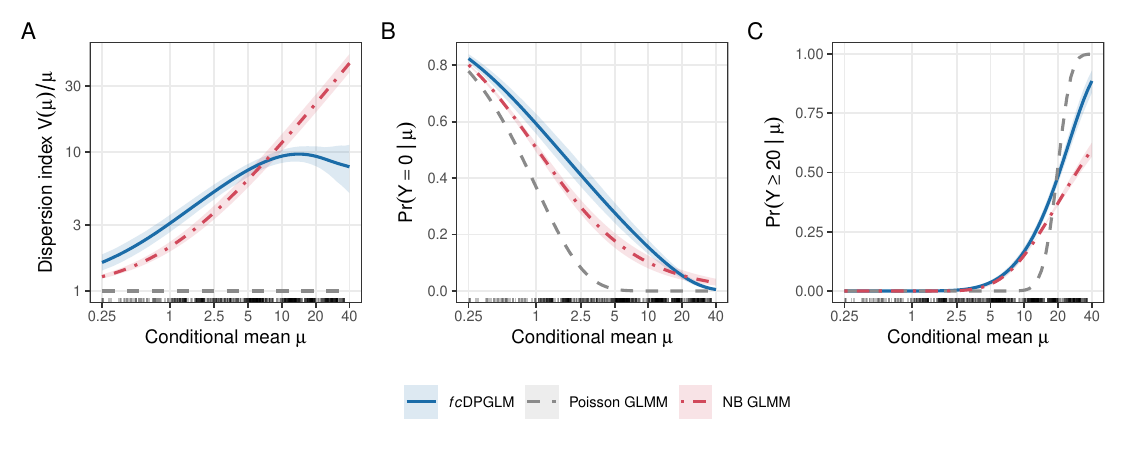}
  \caption{Conditional count distribution as a function of the conditional mean $\mu$ in the seizure study. (A)~Dispersion index $V(\mu)/\mu$, (B)~zero probability $\Pr(Y=0\mid\mu)$ and (C)~exceedance probability $\Pr(Y\geq20\mid\mu)$ under \textit{fc}DPGLM (posterior mean and pointwise $95\%$ credible band) and the negative-binomial GLMM (estimate and $95\%$ interval reflecting uncertainty in its size parameter); the Poisson GLMM gives $V(\mu)/\mu=1$, $\Pr(Y=0\mid\mu)=e^{-\mu}$ and a Poisson tail. A rug along the horizontal axis marks the conditional means of the observed rat-days (posterior means). Tick labels are spaced logarithmically on the horizontal axes and on the vertical axis of (A).}
  \label{fig:case-dispersion}
\end{figure}

Figure~\ref{fig:case-dispersion} shows the source of these differences: the
conditional count distribution implied by each model as a function of the
conditional mean $\mu$. Under \textit{fc}DPGLM the dispersion index (variance-to-mean ratio)
$V(\mu)/\mu$ rises to about $9$ for $\mu$ between $10$ and $20$ and then levels
off, whereas under the negative-binomial GLMM, $V(\mu)/\mu=1+\mu/\phi$ keeps
increasing, to more than twice the \textit{fc}DPGLM value at the largest means. The
negative-binomial model therefore gives too few zeros on low-rate days and, on
high-rate days, spreads the distribution too widely: it puts more mass on
extreme counts but less on the moderately high counts that dominate there, and
so understates $\Pr(Y\geq20\mid\mu)$ (Figure~\ref{fig:case-dispersion}C). For
example, at $\mu=2$ the probability of a seizure-free day is $0.46$ under
\textit{fc}DPGLM against $0.35$ under the NB GLMM, and at
$\mu=20$ the probability of at least $20$ seizures is $0.49$ against $0.37$. The
\textit{fc}DPGLM band widens above $\mu\approx20$, where few rat-days are
observed (values in Table~\ref{tab:supp-dispersion}).

\subsection{Answers to the scientific questions}
\label{sec:case-summary}
The analysis answers the five questions of Section~\ref{sec:application}.
(i)~For a typical rat, the seizure rate rises to a peak of $6.3$ per day around
day~6 of Stage~1 and then declines, whereas in Stage~2 it stays high, near $10$, and
nearly flat. (ii)~Rates are higher in the chronic period on every day of common
follow-up, with the smallest gap at the Stage~1 peak, by $7.1$ $(4.3,11.0)$ seizures per day
on average, a geometric-mean rate ratio of $4.0$. (iii)~The periods differ in
shape, not only in level, consistent with more stable seizure activity in the
chronic period. (iv)~Rats differ substantially after accounting for stage and
time, with a $30.8$-fold range in rates between the most and least affected
rats. (v)~Seizure-free days are more frequent at low rates, and high-burden days
more likely at large means, than the negative-binomial GLMM, because the
dispersion rises with the mean and then levels off. The Poisson and
negative-binomial GLMMs give broadly the same answers to (i) and (ii), so the
mean conclusions are robust to the count distribution, but only
\textit{fc}DPGLM describes seizure-free and high-burden days consistently with
the observed counts.

\section{Discussion}
\label{sec:discussion}

We developed \textit{fc}DPGLM, a semiparametric Bayesian mixed model for
recurrent functional count data that combines GLM-style regression with a
nonparametric count distribution. Regression coefficients and random effects act on the
conditional mean exactly as in a standard count GLMM, while the rest of the
conditional distribution, including its variance, zero probability and tail behavior, is estimated nonparametrically,
without restriction to a parametric family. In the seizure study, the conclusions about the mean, a rise-and-fall pattern
in Stage~1, a higher and flatter trajectory in Stage~2 and substantial
between-rat heterogeneity, were consistent across the considered alternative inference models. The proposed \textit{fc}DPGLM was consistent
with the observed zero frequency, variance and upper tail, whereas the Poisson
GLMM understated all three, and the negative-binomial GLMM underestimated zero
frequencies and overstated the variance and probabilities of extreme
counts. The learned variance function explains why: dispersion rises with the
mean and then levels off, a pattern that no standard parametric count family
produces.

Some limitations remain. Where data are sparse, at high means, the learned reference distribution rests on
little information and, in the tail above the largest observed count, is determined mainly by the prior and the
truncation point $K=\max_iy_i+30$.
A correctly specified parametric model extrapolates into these regions and so
estimates the dispersion more precisely, but it is biased when the family fits poorly. Smoothing the reference distribution, for example by replacing the
independent gamma weights with a mixture of discrete kernels, would borrow
strength across neighboring counts. In addition, all observations share one
reference distribution, so observations with the same conditional mean have
the same distribution; letting the reference distribution depend on
covariates, such as stage, would relax this.

Conditional on the random effects, \textit{fc}DPGLM assumes independence across
repeated observations; richer random-effects structures or serial dependence
could be added when residual temporal correlation matters. Extending the model
to multivariate or marked outcomes would allow seizure type or severity to be
modeled jointly. The seizure study has only $21$ rats, which limits
generalization beyond this experiment, although in simulations with ten and
$21$ rats the intervals for the mean functionals had close to nominal coverage.

In practice, when only mean trajectories and contrasts are of interest and
predictive checks of a negative-binomial mixed model show no lack of fit,
\textit{fc}DPGLM should give essentially the same answers as the parametric
models. The
\textit{fc}DPGLM is preferable when questions also involve zero frequencies,
high-burden days (exceedance probabilities), variability or prediction of new counts, or when predictive checks for inference under parametric models fail, as for the seizure data. In either case, the posterior
on $\alpha$ shows how strongly the data support shrinkage toward the parametric
model obtained as $\alpha\to\infty$ (Proposition~\ref{prop:limit}).

\section*{Acknowledgements}
E.A.\ and P.M.\ were supported by the National Institutes of Health (NIH) under grants 5R01NS133040-02 (E.~Juarez-Colunga, PI) and 2R01~HL094786 (J.~Schildcrout, PI).

\section*{Data Availability Statement}
The seizure data analyzed in this article are not publicly available; they are
available from Marco I.\ Gonzalez upon reasonable request. Code implementing
\textit{fc}DPGLM and reproducing the analysis, with a synthetic data set of the
same structure and the corresponding results, is available at
\url{https://github.com/Entejar/fcDPGLM}.

\bibliographystyle{biom}
\bibliography{references}

\clearpage
\appendix
\setcounter{table}{0}\setcounter{figure}{0}\setcounter{equation}{0}
\renewcommand{\thetable}{S\arabic{table}}
\renewcommand{\thefigure}{S\arabic{figure}}
\renewcommand{\theequation}{S\arabic{equation}}
\section*{Appendices}

Appendix~\ref{sec:supp-data} describes the construction of the rat-day
data. Appendices~\ref{sec:supp-model}--\ref{sec:supp-mcmc} provide the
computational representation of the \fcdpglm{} model, the proofs of the
propositions, the auxiliary-variable model augmentation and the posterior
sampler. Appendices~\ref{sec:supp-sim-design} and~\ref{sec:supp-bands}
provide additional details on the data-generating model and truths for the
simulation study and the pointwise and simultaneous uncertainty bands, and
Appendix~\ref{sec:supp-sim-sens} reports the sensitivity to the prior
specification. Appendices~\ref{sec:supp-mcmc-diagnostics}
and~\ref{sec:supp-additional-results} provide MCMC diagnostics and additional
posterior summaries for the seizure data, Appendix~\ref{sec:supp-simulation} provides additional simulation results,
and Appendix~\ref{sec:supp-ppc} provides additional posterior predictive
checks for the seizure data. 

\section{Data construction and exploratory summaries}
\label{sec:supp-data}

The original recurrent-event data consist of seizure-event records and
tracking intervals for 21 rats recorded in two periods: the early (Stage~1)
and chronic (Stage~2) periods after induced status epilepticus (Section~\ref{sec:application}). The analysis aggregates these records to
the rat-day level. Time is measured within each stage from the start of each
rat's recording in that stage.

For each tracking interval with start and end times measured in seconds
since the start of recording in that stage, the interval is intersected with consecutive
24-hour bins. The amount of overlap with each bin is summed
over all tracking intervals for the same rat and stage. Dividing by 86,400
seconds gives the daily exposure $e_i$. Seizure events are assigned to a
day according to their event end time, and the number of events assigned to
each rat-stage-day combination defines the observed count $y_i$. Rat-stage-days
with positive observed exposure are retained. This construction results in
618 rat-day observations from 21 rats.

\begin{figure}[!ht]
  \centering
  \includegraphics[width=\linewidth]{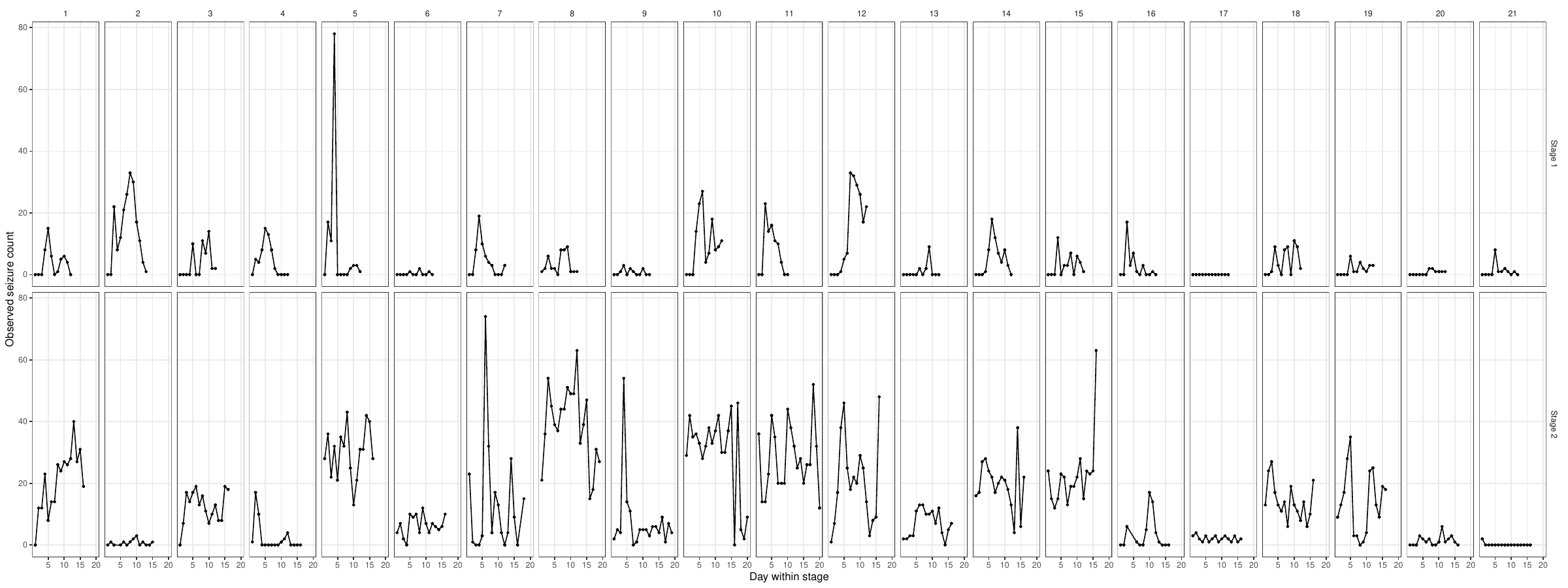}
  \caption{Observed seizure trajectories for each of the 21 rats, shown
  separately by stage.}
  \label{fig:supp-rat-trajectories}
\end{figure}

Exposure is close to one day for most observations but is not exactly one
for every rat-day. Across the 618 observations, the minimum exposure is
approximately $2.55\times10^{-4}$ days, the median is 1 day, and the mean is
0.977 days. The statistical analysis retains all observations and accounts
for exposure through the offset in the conditional mean model. For the
exploratory raw-count figures only (Figure~\ref{fig:rat-eda} and Figure~\ref{fig:supp-rat-trajectories}), we restrict attention to the 598
rat-days with at least 0.95 days of observed exposure, since raw counts from
very short observation intervals are not directly comparable with
full-day counts.

Stage~1 contains 259 rat-days and Stage~2 contains 359; other descriptive
summaries are given in Section~\ref{sec:application}. Seizure type and duration are recorded for each event, but the analysis uses
the total daily count, so it does not distinguish seizures by type or
severity. Figure~\ref{fig:supp-rat-trajectories} shows the observed trajectory of
each rat; the rats differ markedly in both stages.

\section{Computational representation and prior specification}
\label{sec:supp-model}

This appendix describes the computational representation used to implement
the \fcdpglm{} model defined in Section~\ref{sec:model}. We retain its notation. Let $Y_i$ denote the count response, $e_i>0$ the exposure,
$r(i)$ the rat, $d_i\in\{1,2\}$ the stage, and $t_i$ time within
stage. The conditional mean is
\begin{equation}
  \mu_i
  =\E\{Y_i\given d_i,t_i,e_i,a_{r(i)}\}
  =e_i\exp\left(x_i^\top\beta+a_{r(i)}\right),
  \label{eq:supp-mean}
\end{equation}
where
\begin{equation}
  x_i^\top\beta
  =\beta_0+\beta_{\mathrm{stage}}1(d_i=2)
   +B(t_i)^\top\gamma_1 1(d_i=1)
   +B(t_i)^\top\gamma_2 1(d_i=2).
  \label{eq:supp-predictor}
\end{equation}
The basis $B(t)$ is a natural cubic spline basis with three degrees of
freedom, evaluated using time standardized within each stage. We assume
rat-specific random intercepts
\begin{equation}
  a_r\sim\Normal(0,\sigma_a^2),\qquad r=1,\ldots,R.
  \label{eq:supp-a-prior}
\end{equation}
The random-effect standard deviation has prior $\sigma_a\sim$ half-Cauchy$(0,1)$
\citep{gelman2006prior}.
The intercept and stage effect have priors $\Normal(0,10^2)$, and the spline
coefficients of stage $d$, expressed in a basis that is orthonormal over the
observations of that stage, have priors
\begin{equation}
  \gamma_d\given\tau_d\sim\Normal(0,\tau_d^2I),\qquad
  \tau_d\sim\text{half-Cauchy}(0,1),\qquad d=1,2.
  \label{eq:supp-beta-prior}
\end{equation}
As described in Section~\ref{sec:model}, the outcome space is
$\mathbb{N}_0$. We implement posterior inference by way of posterior Markov
chain Monte Carlo (MCMC) simulation. We truncate the count space
$\mathbb{N}_0$ at an upper bound $K$ well beyond the largest observed count
and work on $\mathbb N_{0,K}=\{0,\ldots,K\}$.
For the recurrent seizure analysis we use
$K=\max_i y_i+30=108$. The concentration parameter of the gamma completely random measure (CRM)
\citep{kingman1967completely} has prior $\alpha\sim\GammaDist(1,0.1)$ (shape
and rate). The base distribution
$G_0$ is negative binomial with mean 10.44, equal to the overall
exposure-adjusted empirical seizure rate, and size parameter 0.931, estimated
from a preliminary negative-binomial mixed model. For computation, this
negative-binomial distribution is truncated to $\mathbb N_{0,K}$ and
renormalized so that
$
  G_0(k)\geq 0, \ \ \sum_{k \in \mathbb N_{0,K}} G_0(k)=1.
$
On $\mathbb N_{0,K}$, the gamma CRM is represented by independent random
masses
\begin{equation}
  \lambda=\sum_{k \in \mathbb N_{0,K}} w_k\delta_k,
  \qquad
  w_k\stackrel{\mathrm{ind}}{\sim}
  \GammaDist\{\alpha G_0(k),1\}.
  \label{eq:supp-crm}
\end{equation}
This is the computationally truncated representation of the gamma CRM on
$\mathbb{N}_0$ in Section~\ref{sec:crm}. For a tilting parameter $\theta_i$, define
\begin{equation}
  b_i(\theta_i;w)
  =\log\left\{\sum_{k \in \mathbb N_{0,K}} w_k\exp(\theta_i k)\right\}.
  \label{eq:supp-b}
\end{equation}
The likelihood contribution for an observed count $y_i$ is
\begin{equation}
  p(y_i\mid\theta_i,w)
  =\exp\{\theta_i y_i-b_i(\theta_i;w)\}w_{y_i}.
  \label{eq:supp-likelihood}
\end{equation}
The tilting parameter is not a free parameter and thus is not sampled; for
each observation it is determined by the mean constraint
\begin{equation}
  \frac{\sum_{k \in \mathbb N_{0,K}} k w_k\exp(\theta_i k)}
       {\sum_{k \in \mathbb N_{0,K}} w_k\exp(\theta_i k)}
  =\mu_i.
  \label{eq:supp-theta-inversion}
\end{equation}

\section{\texorpdfstring{Proofs of Propositions~\ref{prop:limit}--\ref{prop:ident} and Remark~\ref{rem:score}}{Proofs of Propositions 1--2 and Remark 1}}
\label{sec:supp-proofs}
Throughout, the count space is $\mathbb N_{0,K}$ and, for a probability mass
function $\pi$ on it that is not concentrated at a single point,
$Q^\pi_\mu(k)=\pi(k)e^{\theta k}/\sum_\ell\pi(\ell)e^{\theta\ell}$ with $\theta=\theta(\mu,\pi)$
solving
\begin{equation}
  h(\theta,\pi)\equiv\frac{\sum_kk\,\pi(k)e^{\theta k}}{\sum_k\pi(k)e^{\theta k}}=\mu .
  \label{eq:supp-h}
\end{equation}
Since $\partial h/\partial\theta$ is the variance of $Q^\pi_\mu$, which is
positive, $h$ is strictly increasing in $\theta$. If $\pi(0)>0$ and
$\pi(K)>0$, $h\to0$ as $\theta\to-\infty$ and $h\to K$ as $\theta\to\infty$, so
for $0<\mu<K$ \eqref{eq:supp-h} has a unique solution.

\paragraph{Proof of Proposition~\ref{prop:limit}.}
Under the truncated prior, $w_k\sim\GammaDist\{\alpha G_0(k),1\}$
independently, so $f_0=w/\sum_kw_k\sim\mathrm{Dirichlet}\{\alpha G_0(0),\ldots,\alpha G_0(K)\}$
\citep{ferguson1973bayesian},
with $E\{f_0(k)\}=G_0(k)$ and $\operatorname{Var}\{f_0(k)\}=G_0(k)\{1-G_0(k)\}/(\alpha+1)$.
Hence $f_0\to G_0$ in probability as $\alpha\to\infty$ by Chebyshev's
inequality. The function $h$ is continuously differentiable in
$(\theta,\pi)$ on the interior of the simplex, which contains $G_0$, with
$\partial h/\partial\theta>0$; by the implicit function theorem
$\theta(\mu,\pi)$ is continuous at $\pi=G_0$, and so is
$Q^\pi_\mu(k)$ for every $k$. By the continuous mapping theorem,
$p_i=Q^{f_0}_{\mu_i}\to Q^{G_0}_{\mu_i}$ in probability for each $i$, and the
likelihood $\prod_ip_i(y_i)$, a continuous function of finitely many such
terms, converges to $\prod_iQ^{G_0}_{\mu_i}(y_i)$.
If $G_0(k)\propto g(k)q^k$ on $\mathbb N_{0,K}$, then
$G_0(k)e^{\theta k}\propto g(k)(q e^{\theta})^k$, the member of the same
family with parameter $q e^\theta$, truncated to $\mathbb N_{0,K}$; the
mean constraint selects the member with mean $\mu$. For the Poisson family
$g(k)=1/k!$, and for the negative-binomial family with size $r$,
$g(k)=\Gamma(k+r)/\{k!\,\Gamma(r)\}$ and $q=\bar \mu/(\bar \mu+r)$, where
$\bar \mu$ is the mean of $G_0$; tilting changes $q$ but not $r$.
\hfill$\square$

\paragraph{Proof of Proposition~\ref{prop:ident}.}
With $w'_k=c\,e^{tk}w_k$ and $\theta'_i=\theta_i-t$,
\[
  \frac{w'_ke^{\theta'_ik}}{\sum_\ell w'_\ell e^{\theta'_i\ell}}
  =\frac{c\,w_ke^{\theta_ik}}{c\sum_\ell w_\ell e^{\theta_i\ell}}=p_i(k),
\]
so the tilt of $w'$ with parameter $\theta'_i$ is $p_i$, whose mean is $\mu_i$.
By uniqueness of the solution of the mean constraint, $\theta'_i$ is the
derived tilting parameter under $w'$, and the conditional distributions,
hence the likelihood, are unchanged. The same argument with $\mu_i$ replaced by
any $\mu\in(0,K)$ gives $Q^{f_0'}_\mu=Q^{f_0}_\mu$, where $f_0'$ is the normalized
$w'$. \hfill$\square$

\paragraph{Derivation of Remark~\ref{rem:score}.}
With the weights fixed, $\log p_i(y_i)=\log w_{y_i}+\theta_iy_i-b(\theta_i;\lambda)$,
so $\partial\log p_i(y_i)/\partial\theta_i=y_i-b'(\theta_i;\lambda)=y_i-\mu_i$.
Differentiating the mean constraint $b'(\theta_i;\lambda)=\mu_i$ with respect to
$\mu_i$ gives $b''(\theta_i;\lambda)\,\partial\theta_i/\partial \mu_i=1$, so
$\partial\theta_i/\partial \mu_i=1/V_i$, and $\partial \mu_i/\partial\eta_i=\mu_i$
under the log link. The chain rule gives the score
$(y_i-\mu_i)\mu_i/V_i$. Its square has expectation
$E\{(Y_i-\mu_i)^2\}\mu_i^2/V_i^2=\mu_i^2/V_i$ under $p_i$. Finally, the score has
mean zero under any distribution of $Y_i$ with mean $\mu_i$, which gives the
unbiasedness of the estimating equations noted after the remark.
\hfill$\square$

The proof of Proposition~\ref{prop:mass} is given in Appendix~\ref{sec:supp-mass}. 

\section{Model augmentation with auxiliary variable}
\label{sec:supp-augmentation}

For the CRM update, define
\begin{equation}
  T_i(w)=\sum_{k \in \mathbb N_{0,K}} w_k\exp(\theta_i k).
  \label{eq:supp-T}
\end{equation}
Using
\begin{equation}
  \frac{1}{T_i(w)}
  =\int_0^\infty \exp\{-u_iT_i(w)\}\,\dd u_i,
  \label{eq:supp-u-identity}
\end{equation}
we introduce auxiliary variables satisfying
\begin{equation}
  u_i\given w,\theta_i\sim\GammaDist\{1,T_i(w)\},
  \label{eq:supp-u-conditional}
\end{equation}
where the second argument is the rate. Let
\begin{equation}
  \psi(k)=\sum_{i=1}^n u_i\exp(\theta_i k),
  \qquad
  n_k=\sum_{i=1}^n1(y_i=k),
  \label{eq:supp-psi}
\end{equation}
and define
\begin{equation}
  c_k=\alpha G_0(k)+n_k.
  \label{eq:supp-shape}
\end{equation}
Holding the tilting parameters fixed, the augmented kernel for the CRM
weights is gamma, giving
\begin{equation}
  w_k\given u,\theta
  \sim\GammaDist\{c_k,1+\psi(k)\},
  \qquad k \in \mathbb N_{0,K}.
  \label{eq:supp-w-conditional}
\end{equation}
Equation~\eqref{eq:supp-u-conditional} is an exact full conditional. If the
$\theta_i$ were fixed parameters, \eqref{eq:supp-w-conditional} would also be
an exact full conditional. In \fcdpglm{}, however, $\theta_i$ changes
when the CRM weights change because it is determined by
\eqref{eq:supp-theta-inversion}. We therefore use
\eqref{eq:supp-w-conditional} as an informed proposal and apply the
Metropolis--Hastings correction described in Appendix~\ref{sec:supp-crm}.

\section{Markov chain Monte Carlo posterior simulation}
\label{sec:supp-mcmc}

We describe the transition probabilities of the MCMC sampler for \fcdpglm{}.
All updates leave the posterior under the computational truncation invariant.
One iteration consists of the following steps.
\begin{enumerate}[label=\arabic*.]
  \item Joint update of $(\beta,a)$ by a Fisher-preconditioned
        Metropolis-adjusted Langevin step (Appendix~\ref{sec:supp-mala}).
  \item Conjugate draws of $\sigma_a$ and of the spline scales $\tau_1,\tau_2$
        (Appendix~\ref{sec:supp-scales}).
  \item Gibbs draw of the auxiliary variables $u$
        (Appendix~\ref{sec:supp-u}).
  \item Gibbs draw of the total CRM mass (Appendix~\ref{sec:supp-mass}).
  \item Update of the CRM weights in four random blocks, each by the conjugate
        proposal with a Metropolis--Hastings correction
        (Appendix~\ref{sec:supp-crm}).
  \item Slice-sampling update of $\alpha$ from its full conditional
        (Appendix~\ref{sec:supp-alpha}).
\end{enumerate}
Throughout, $\theta$ is recomputed from the mean constraint
\eqref{eq:supp-theta-inversion} whenever $(\beta,a)$ or the weights change.

\subsection{Joint update of the regression coefficients and rat effects}
\label{sec:supp-mala}

Write $\xi=(\beta^\top,a^\top)^\top$, $\eta_i=x_i^\top\beta+a_{r(i)}$ and
$\eta=Z\xi$ with $Z=[X,D]$, where $D$ is the rat-indicator matrix. With the
CRM weights held fixed, each $\theta_i$ depends on $\xi$ only through $\eta_i$.
Implicit differentiation of $b_i'(\theta_i;w)=\mu_i$ gives
$\partial\theta_i/\partial \mu_i=1/V_i$, where $V_i=b_i''(\theta_i;w)$ is the
tilted variance, and hence
\[
  \frac{\partial\{\theta_iy_i-b_i(\theta_i;w)\}}{\partial\eta_i}
  =\frac{(y_i-\mu_i)\,\mu_i}{V_i}.
\]
The gradient of the log posterior is therefore $g(\xi)=Z^\top\kappa-\Lambda(\xi-\xi_0)$,
with $\kappa_i=(y_i-\mu_i)\mu_i/V_i$ and $\Lambda$ and $\xi_0$ the prior precision
and mean of $\xi$ given the current scales. We use the Fisher information plus
the prior precision,
\[
  H(\xi)=Z^\top\mathrm{diag}(\mu_i^2/V_i)\,Z+\Lambda,
\]
as a position-dependent preconditioner \citep{girolami2011riemann} and
propose
\[
  \xi^\star\sim\Normal\Bigl\{\xi+\tfrac{\epsilon^2}{2}H(\xi)^{-1}g(\xi),\;
  \epsilon^2H(\xi)^{-1}\Bigr\}.
\]
The proposal is accepted with the Metropolis--Hastings ratio that uses the
proposal density in both directions, including the determinant of
$H(\xi^\star)$, so the update is exact although the preconditioner depends on
the state. The step size $\epsilon$ is adapted toward an acceptance rate of
$0.574$ during the adaptation phase and fixed afterwards. To use a single prior
scale for each stage's trajectory, the spline columns of each stage are
orthonormalized over that stage's observations before fitting, and draws are
transformed back to the original basis for reporting.

\subsection{Conjugate updates of the variance scales}
\label{sec:supp-scales}

Each half-Cauchy$(0,A)$ scale $s$ with $m$ normal coefficients
$b_1,\ldots,b_m\sim N(0,s^2)$ is updated through the inverse-gamma mixture
representation of \citet{makalic2016simple},
$s^2\given\nu\sim\mathrm{IG}(1/2,1/\nu)$, $\nu\sim\mathrm{IG}(1/2,1/A^2)$:
\[
  s^2\given b,\nu\sim\mathrm{IG}\Bigl(\tfrac{m+1}{2},\,\tfrac1\nu+\tfrac12\textstyle\sum_jb_j^2\Bigr),
  \qquad
  \nu\given s^2\sim\mathrm{IG}\bigl(1,\,1/A^2+1/s^2\bigr).
\]
This is applied to $\sigma_a$ (with the $R$ rat effects) and to $\tau_1$ and
$\tau_2$ (with the orthonormalized spline coefficients of each stage), all with
$A=1$.

\subsection{Auxiliary-variable update}
\label{sec:supp-u}

Given the current weights and tilting parameters, the $u_i$ are drawn
independently from their full conditional distribution. Numerically we use
\begin{equation}
  \log u_i=\log E_i-b_i(\theta_i;w),
  \qquad E_i\sim\operatorname{Exponential}(1),
  \label{eq:supp-u-draw}
\end{equation}
which is equivalent to $u_i\sim\operatorname{Exponential}\{T_i(w)\}$.

\subsection{Gibbs draw of the total CRM mass}
\label{sec:supp-mass}

The likelihood depends on the weights only through their relative sizes, so
the total mass $M=\sum_kw_k$ is a weakly identified direction that the other
updates move only slowly. Proposition~\ref{prop:mass} gives a Gibbs
update; we prove it here.

\paragraph{Proof of Proposition~\ref{prop:mass}.}
(i)~The tilting parameter $\theta_i$ solves
$\sum_kk\,w_ke^{\theta_ik}/\sum_kw_ke^{\theta_ik}=\mu_i$, and both sums scale
by $M$ when $w=Mf_0$, so $\theta_i=\theta(\mu_i,f_0)$.
(ii)~Let $L=K+1$ be the number of support points and write $\phi$ for
$(\beta,a,\alpha)$ and the remaining parameters. Conditionally on $\phi$, the
augmented posterior density of $(w,u)$ is proportional to
\[
  \prod_{k=0}^{K}w_k^{\alpha G_0(k)-1}e^{-w_k}
  \;\prod_{i=1}^n w_{y_i}e^{\theta_iy_i}\exp\{-u_iT_i(w)\},
\]
where $G_0$ is the truncated and renormalized base distribution, so that
$\sum_kG_0(k)=1$. Change variables from $(w,u)$ to $(M,f_0(0),\ldots,f_0(K-1),v)$
with $w=Mf_0$ and $v=Mu$; the Jacobian is $M^{L-1}\cdot M^{-n}$. Since
$\prod_kw_k^{\alpha G_0(k)-1}=M^{\alpha-L}\prod_kf_0(k)^{\alpha G_0(k)-1}$,
$\sum_kw_k=M$, $\prod_iw_{y_i}=M^n\prod_if_0(y_i)$, $u_iT_i(w)=v_iT_i(f_0)$,
and $\theta_i$ depends only on $f_0$ by~(i), the transformed density is
\[
  M^{\alpha-L}\,e^{-M}\,M^{n}\,M^{L-1}M^{-n}\;
  g(f_0,v,\phi)
  =M^{\alpha-1}e^{-M}\,g(f_0,v,\phi),
\]
for a function $g$ not involving $M$. Hence $M$ given $(f_0,v,\phi)$ and the
data is $\GammaDist(\alpha,1)$. Drawing $M^\star$ from this conditional and
mapping back gives $w^\star=M^\star f_0=(M^\star/M)w$ and
$u^\star=v/M^\star=(M/M^\star)u$. \hfill$\square$

\medskip\noindent
Because $g$ does not involve $M$, the conditional of $\alpha$ given
$(f_0,v)$ depends on the weights only through $f_0$: the data inform
$\alpha$ only through the normalized weights. A Metropolis--Hastings move
that changes $\alpha$ while holding the unnormalized weights fixed therefore
conflicts with the $\GammaDist(\alpha,1)$ law of $M$ and is accepted only for
small changes of $\alpha$; the sampler instead draws $M$ from its conditional and updates
$\alpha$ by slice sampling (Appendix~\ref{sec:supp-alpha}).

\subsection{Block updates of the CRM weights}
\label{sec:supp-crm}

The support $\mathbb N_{0,K}$ is split at random into four blocks at each
iteration, and the blocks are updated in turn. For the full support, the
conjugate proposal is
\begin{equation}
  w_k^\star\sim\GammaDist\{c_k,1+\psi(k)\},
  \qquad k \in \mathbb N_{0,K},
  \label{eq:supp-crm-proposal}
\end{equation}
after which $\theta^\star$ is recomputed from the mean constraint and the
proposal is accepted with probability $\min(1,R_{\mathrm{CRM}})$, where, with
$\psi^\star(k)=\sum_iu_i\exp(\theta_i^\star k)$ and
$T_i(w;\theta)=\sum_kw_k\exp(\theta_ik)$,
\begin{align}
  \log R_{\mathrm{CRM}}={}&
  \sum_{i=1}^n(\theta_i^\star-\theta_i)y_i
  +\sum_{i=1}^n u_i\big[
       T_i(w^\star;\theta)+T_i(w;\theta)
       -T_i(w^\star;\theta^\star)-T_i(w;\theta^\star)
     \big] \nonumber\\
  &+\sum_{k \in \mathbb N_{0,K}} c_k
     \left[\log\{1+\psi^\star(k)\}-\log\{1+\psi(k)\}\right].
  \label{eq:supp-crm-ratio}
\end{align}
The gamma kernels cancel because the proposal is the conditional gamma kernel
evaluated at the current tilting parameters; the correction is needed only
because the tilting parameters change with the weights. For a block $B$, only
$w_k$, $k\in B$, are proposed from \eqref{eq:supp-crm-proposal}, and the
ratio becomes
\[
  \log R_B
  =\sum_i(\theta_i^\star-\theta_i)y_i
  +\sum_iu_i\bigl\{T_i(\tilde w;\theta)-T_i(\tilde w;\theta^\star)\bigr\}
  +\sum_{k\in B}c_k\bigl[\log\{1+\psi^\star(k)\}-\log\{1+\psi(k)\}\bigr],
\]
with $\tilde w_k=w_k^\star+w_k$ for $k\in B$ and $\tilde w_k=w_k$ otherwise; when
$B$ is the whole support this equals \eqref{eq:supp-crm-ratio}. Changing a
block moves the tilting parameters less than changing all weights at once,
which raises the acceptance rate.

\subsection{Update of the concentration parameter}
\label{sec:supp-alpha}

Given the weights, the conditional of $\alpha$ is
\[
  p(\alpha\given w)\propto p(\alpha)\prod_{k\in\mathbb N_{0,K}}
  \frac{w_k^{\alpha G_0(k)}}{\Gamma\{\alpha G_0(k)\}},
\]
which we sample on the log scale by slice sampling, using the stepping-out
and shrinkage procedures of \citet{neal2003slice}. This conditional depends on
$\log w_k$ for every $k$, including counts that are not observed, whose
weights can fall below the smallest positive double-precision number because
their gamma shapes $c_k=\alpha G_0(k)$ are tiny. We therefore track the
weights on the logarithmic scale, drawing each as
$\log G_1+\log U/c_k-\log\{1+\psi(k)\}$ with $G_1\sim\GammaDist(c_k+1,1)$ and
$U\sim\operatorname{Uniform}(0,1)$.

\section{Simulation design and data-generating model}
\label{sec:supp-sim-design}

\subsection{Data-generating model}
\label{sec:supp-sim-truth}

The simulation truth was estimated once from the seizure data by the
semiparametric maximum-likelihood estimator of the SPGLM
\citep{rathouz2009generalized}, computed with the \texttt{gldrm} package
\citep{wurm2018semiparametric},
with the mean model of Section~\ref{sec:functional-model} and the $21$ rats entered as fixed
effects constrained to sum to zero. We treat the resulting estimates as the
true values $\beta^\star$, $a^\star_r$ and $f_0^\star$.
Table~\ref{tab:supp-sim-beta} gives the regression coefficients $\beta^\star$,
in the parameterization of \eqref{eq:supp-predictor} with the natural-spline
basis of the original (not orthonormalized) time scale. Both scenarios, negative binomial and
semiparametric, use the same coefficients, so they differ only in the
conditional distribution of the counts given their means.
\begin{table}[!ht]
  \centering
  \caption{True regression coefficients. In the null-shape truth, the
  intercept and Stage~1 coefficients are replaced so that the two stages
  differ in level only; the Stage~2 curve is unchanged
  ($\beta_0+\beta_{\mathrm{stage}}=2.12$ in both truths, up to rounding, and
  identical Stage~2 spline coefficients).}
  \label{tab:supp-sim-beta}
  \small
  \begin{tabular}{lrr}
    \toprule
    Coefficient & Shape-difference truth & Null-shape truth \\
    \midrule
    Intercept $\beta_0$                  & $-1.146$ & $0.714$ \\
    Stage effect $\beta_{\mathrm{stage}}$ & $3.268$ & $1.407$ \\
    Stage~1 spline 1                      & $0.829$  & $0.242$ \\
    Stage~1 spline 2                      & $4.815$  & $0.355$ \\
    Stage~1 spline 3                      & $-0.384$ & $-0.022$ \\
    Stage~2 spline 1                      & $0.136$  & $0.136$ \\
    Stage~2 spline 2                      & $0.172$  & $0.172$ \\
    Stage~2 spline 3                      & $-0.424$ & $-0.424$ \\
    \bottomrule
  \end{tabular}
\end{table}
Under the shape-difference truth the average difference is
$\bar\Delta=6.94$ seizures per day and the average log rate ratio is $1.40$;
under the null-shape truth the average log rate ratio is also $1.40$ and the
shape contrast is $S(t)=0$ up to a projection error of at most $0.0012$
(construction in Appendix~\ref{sec:supp-sim-null}).

The estimated rat effects, centered to sum to zero and used as the true values $a^\star_r$, are given in
Table~\ref{tab:supp-sim-a}. They are not
used as fixed truths for particular rats. Instead, in each replicate the
$R$ rat effects ($R=21$ or $10$) are drawn with replacement from these $21$
values, independently of which rats' designs are used. The true
random-effect distribution is therefore the discrete distribution placing
mass $1/21$ on each value in Table~\ref{tab:supp-sim-a}, with standard
deviation $1.01$.
\begin{table}[!ht]
  \centering
  \caption{True rat effects $a^\star_r$, resampled with replacement in each
  replicate of the simulation.}
  \label{tab:supp-sim-a}
  \small
  \setlength{\tabcolsep}{4pt}
  \begin{tabular}{lrrrrrrr}
    \toprule
    Rat & 1 & 2 & 3 & 4 & 5 & 6 & 7 \\
    $a^\star_r$ & $0.647$ & $0.221$ & $0.212$ & $-0.588$ & $1.120$ & $-0.702$ & $0.282$ \\
    \midrule
    Rat & 8 & 9 & 10 & 11 & 12 & 13 & 14 \\
    $a^\star_r$ & $1.296$ & $-0.414$ & $1.151$ & $1.040$ & $0.917$ & $-0.502$ & $0.640$ \\
    \midrule
    Rat & 15 & 16 & 17 & 18 & 19 & 20 & 21 \\
    $a^\star_r$ & $0.698$ & $-0.571$ & $-1.878$ & $0.351$ & $0.181$ & $-1.829$ & $-2.272$ \\
    \bottomrule
  \end{tabular}
\end{table}

The two conditional count distributions are as follows. In the negative
binomial scenario, $Y_i$ is negative binomial with mean $\mu_i$ and size
$0.982$ (variance $\mu_i+\mu_i^2/0.982$), the size estimated by a
negative-binomial GLM with the same mean model. In the semiparametric
scenario, $Y_i\sim Q^{f_0^\star}_{\mu_i}$, the exponential tilt, to mean $\mu_i$,
of the \texttt{gldrm} estimate $f_0^\star$ of the reference distribution,
which is supported on the $54$ distinct observed counts between $0$ and $78$;
Figure~\ref{fig:supp-sim-truth}(B) shows both distributions at three means. In both
scenarios $\mu_i=e_i\exp(x_i^\top\beta^\star+a_{r(i)})$, where $a_{r(i)}$ is the
rat effect drawn for that replicate and the stages, days and exposures are
those of the rats used in that replicate.

\subsection{Construction of the null-shape truth}
\label{sec:supp-sim-null}

The null-shape settings check that the simultaneous band for $S(t)$ has the
right false-positive rate. They need mean curves for which the two stages
differ in level only, so that $S(t)=0$, while everything else is as in the
shape-difference truth: the Stage~2 curve, the size of the stage difference,
the rat effects, the count distribution and the design.

\paragraph{Notation.} For a typical rat ($a=0$) with one day of exposure, the
log mean curves on the common grid $\mathcal T=\{1,\ldots,13\}$ are
\[
  \eta_1(t)=\beta_0+B(t)^\top\gamma_1,\qquad
  \eta_2(t)=\beta_0+\beta_{\mathrm{stage}}+B(t)^\top\gamma_2,
\]
where $B(t)$ is the natural-spline basis evaluated at time standardized
within the stage. The shape contrast is
$S(t)=\{\eta_2(t)-\bar\eta_2\}-\{\eta_1(t)-\bar\eta_1\}$, where $\bar\eta_1$
and $\bar\eta_2$ are the averages of $\eta_1(t)$ and $\eta_2(t)$ over
$\mathcal T$, so $S\equiv0$ exactly when $\eta_2(t)-\eta_1(t)$
is constant over $\mathcal T$. The average log rate ratio is
$c_0=\bar\eta_2-\bar\eta_1=1.404$ under the shape-difference truth.

\paragraph{Requirements.} We want new coefficients
$(\tilde\beta_0,\tilde\gamma_1,\tilde\beta_{\mathrm{stage}})$ such that
(i)~the Stage~2 curve is unchanged, $\tilde\eta_2=\eta_2$, and
(ii)~$\tilde\eta_1(t)=\eta_2(t)-c_0$ on $\mathcal T$. Then the stages differ by
the constant $c_0$, which is the average log rate ratio of the
shape-difference truth, so the size of the stage difference on the rate-ratio
scale is preserved.

\paragraph{Construction.} The new Stage~1 curve must have the form
$\tilde\beta_0+B(t)^\top\tilde\gamma_1$, a family with four parameters,
whereas the target $\eta_2(t)-c_0$ is a function on the $13$ days of the
common follow-up period $\mathcal T$, so exact equality is not guaranteed. We
therefore take the least-squares fit on $\mathcal T$,
\[
  (\tilde\beta_0,\tilde\gamma_1)
  =\arg\min_{\beta_0,\gamma}\sum_{t\in\mathcal T}
  \bigl\{\eta_2(t)-c_0-\beta_0-B(t)^\top\gamma\bigr\}^2 ,
\]
which is a single linear regression of $\eta_2-c_0$ on an intercept and the
three Stage~1 spline columns over $\mathcal T$. The Stage~2 spline
coefficients $\gamma_2$ are kept, and the stage effect is reset to
\[
  \tilde\beta_{\mathrm{stage}}=\beta_0+\beta_{\mathrm{stage}}-\tilde\beta_0,
\]
so that $\tilde\beta_0+\tilde\beta_{\mathrm{stage}}=\beta_0+\beta_{\mathrm{stage}}$
and the Stage~2 curve is exactly unchanged. Only the mean coefficients change
(Table~\ref{tab:supp-sim-beta}).

\paragraph{Properties.} Let
$r(t)=\eta_2(t)-c_0-\tilde\eta_1(t)$ be the least-squares residual. Because
the regression includes an intercept, $\sum_{t\in\mathcal T}r(t)=0$. Hence:
(i)~$\tilde\eta_2=\eta_2$ exactly; (ii)~the average log rate ratio is exactly
$c_0=1.404$ in both truths, since $\bar{\tilde\eta}_1=\bar\eta_2-c_0$; and
(iii)~the shape contrast of the null-shape truth equals the projection
residual, $\tilde S(t)=r(t)$. The null is therefore exact up to the
projection error, $\max_t|\tilde S(t)|=0.0012$ on the log scale, against
$\max_t|S(t)|=1.82$ under the shape-difference truth. The error is small
because the Stage~1 and Stage~2 spline spaces, which differ only through the
stage-specific standardization of time, nearly coincide on days $1$--$13$;
the two are not identical, because the knots, fixed on the standardized
scale, fall on different days in the two stages.

\paragraph{Direction of the projection.} We project the Stage~2 shape onto the
Stage~1 spline space rather than the reverse. Making Stage~2 copy Stage~1 is
not feasible: the Stage~2 basis is standardized over days $1$--$21$ and cannot
reproduce the Stage~1 peak on day~$6$, and its extrapolation to days
$14$--$21$, where only Stage~2 is observed, produces implausibly large means.

\paragraph{Resulting curves.} Under the null-shape truth, the Stage~1 curve is
nearly flat, between $2.0$ and $2.7$ seizures per day, and the Stage~2 curve
is unchanged (Table~\ref{tab:supp-sim-null-curves}). The average log rate
ratio is $1.40$, as under the shape-difference truth, and the average
difference is $\bar\Delta=7.56$ seizures per day ($6.94$ under the
shape-difference truth). The true values of quantities averaged over the
rat-days, such as the variance and $\Pr(Y\geq20)$, are computed separately
for each truth.
\begin{table}[!ht]
  \centering
  \caption{True stage-specific mean curves $m_d(t)$ (seizures per day) for a
  typical rat with one day of exposure under the shape-difference and
  null-shape truths. The Stage~2 curve is the same in both.}
  \label{tab:supp-sim-null-curves}
  \small
  \setlength{\tabcolsep}{3.5pt}
  \resizebox{\linewidth}{!}{%
  \begin{tabular}{lrrrrrrrrrrrrr}
    \toprule
    Day $t$ & 1 & 2 & 3 & 4 & 5 & 6 & 7 & 8 & 9 & 10 & 11 & 12 & 13 \\
    \midrule
    Stage~2 (in both truths)    & 8.3 & 8.8 & 9.3 & 9.8 & 10.2 & 10.5 & 10.7 & 10.8 & 10.8 & 10.7 & 10.5 & 10.2 & 9.8 \\
    Stage~1, shape difference   & 0.3 & 0.9 & 2.0 & 3.9 & 5.6 & 6.0 & 5.3 & 4.2 & 3.3 & 2.7 & 2.3 & 2.0 & 1.8 \\
    Stage~1, null shape         & 2.0 & 2.2 & 2.3 & 2.4 & 2.5 & 2.6 & 2.6 & 2.7 & 2.7 & 2.6 & 2.6 & 2.5 & 2.4 \\
    \bottomrule
  \end{tabular}}
\end{table}

\section{Pointwise and simultaneous uncertainty bands}
\label{sec:supp-bands}

This appendix describes how the uncertainty bands for the functional estimands
are formed in the recurrent seizure analysis and in the simulation study.

\paragraph{Draws of a functional.} Let $\vartheta(t)$ denote one of the
curves $m_1(t)$, $m_2(t)$, $\Delta(t)$, $RR(t)$ or $S(t)$ of Section~\ref{sec:inference},
for a typical rat ($a=0$) with one day of exposure, evaluated on the common
day grid $\mathcal T=\{1,\ldots,13\}$. Each retained posterior draw of the
regression coefficients gives one draw $\vartheta^{(b)}(t)$, $t\in\mathcal T$,
by evaluating the curve at that draw, $b=1,\ldots,B$. We use $B=20{,}000$ in
Section~\ref{sec:realdata} and $B=2{,}000$ for each simulated data set.

\paragraph{Pointwise bands.} For each day $t$, the pointwise $95\%$ credible
interval is the equal-tailed interval formed by the $2.5\%$ and $97.5\%$
sample quantiles of $\{\vartheta^{(b)}(t)\}_{b=1}^B$.

\paragraph{Simultaneous (sup-$t$) bands.} Let
$\bar\vartheta(t)=B^{-1}\sum_b\vartheta^{(b)}(t)$ and
$\hat s(t)=\{(B-1)^{-1}\sum_b(\vartheta^{(b)}(t)-\bar\vartheta(t))^2\}^{1/2}$
be the posterior mean and standard deviation at day $t$. For each draw
compute the largest standardized deviation across the grid,
\[
  z^{(b)}=\max_{t\in\mathcal T}
  \frac{|\vartheta^{(b)}(t)-\bar\vartheta(t)|}{\hat s(t)},
\]
and let $c$ be the $95\%$ sample quantile of $\{z^{(b)}\}_{b=1}^B$. The
simultaneous $95\%$ band is
\[
  \bar\vartheta(t)\pm c\,\hat s(t),\qquad t\in\mathcal T.
\]
By construction, a fraction $0.95$ of the draws lies entirely inside the band
on all of $\mathcal T$, so the band is a joint statement about the days of
the common follow-up period, and the constant $c$ exceeds the pointwise value
(about $1.96$ for a normal posterior). The bands refer to the days in
$\mathcal T$, not to a continuous time interval.

\paragraph{Use of the bands.} In Section~\ref{sec:realdata}, the simultaneous
bands for $\Delta(t)$ and $S(t)$ are those of Figure~\ref{fig:case-functionals}. In
the simulation study, a difference in temporal shape is declared when the
simultaneous band for $S(t)$ excludes zero on at least one day, and the
simultaneous-band coverage of a curve is the proportion of the $200$
replicates in which its band contains the true curve on all of $\mathcal T$.

\paragraph{Parametric competitors.} For the Poisson and NB mixed models,
$B=2{,}000$ draws of the fixed effects are generated from the asymptotic
normal distribution with the estimated covariance matrix from
\texttt{glmmTMB} \citep{brooks2017glmmtmb}, with the NB size parameter held at
its estimate. Each draw is
mapped to the curves $\vartheta^{(b)}(t)$ in the same way as a posterior draw
of the regression coefficients, and the pointwise and simultaneous bands are
then formed from these draws exactly as above.

\paragraph{Distributional functionals.} For the dispersion index, the zero
probability and the exceedance probability, we report pointwise $95\%$
intervals only.

\section{Sensitivity to the prior specification}
\label{sec:supp-sim-sens}

Table~\ref{tab:supp-sim-sens} compares the prior specification of Section~\ref{sec:functional-model} with a baseline ($\alpha=1$, $\sigma_a$ fixed at the NB mixed-model
estimate, independent $N(0,2.5^2)$ priors on all regression coefficients) and
with each single change from this baseline. All variants were fitted to the
same first $40$ simulated data sets of each $21$-rat shape-difference setting,
so the comparisons are paired. The $N(0,2.5^2)$ prior shrinks the Stage~1
spline coefficients (the largest is about $4.8$) and so biases the shape
contrast; the hierarchical prior and the $N(0,5^2)$ and $N(0,10^2)$ priors
reduce this bias by about two thirds under the NB scenario and by a quarter to
a half under the semiparametric scenario. Sampling $\sigma_a$ slightly improves
the coverage of the count-scale contrasts, mainly under the semiparametric
scenario. The combined specification gives the best coverage of these
contrasts under the semiparametric scenario, at the cost of a slightly larger
bias of $S(t)$ than the $N(0,5^2)$ and $N(0,10^2)$ priors.

\begin{table}[!ht]
  \centering
  \caption{Sensitivity of \fcdpglm\ to the prior specification: integrated
  absolute bias and simultaneous-band coverage (Sim) of $S(t)$, mean pointwise
  coverage of $\Delta(t)$, coverage of the average log rate ratio and of
  $\bar\Delta$. $R=21$ rats and shape-difference mean curves;
  the same $40$ data sets per scenario for every variant.}
  \label{tab:supp-sim-sens}
  \small
  \begin{tabular}{llrrrrr}
    \toprule
    & & \multicolumn{2}{c}{$S(t)$} & $\Delta(t)$ & Avg.\ log RR & $\bar\Delta$ \\
    \cmidrule(lr){3-4}
    Scenario & Specification & $|$Bias$|$ & Sim & Cov & Cov & Cov \\
    \midrule
    NB & Baseline& 0.061 & 0.900 & 0.944 & 0.950 & 0.975 \\
       & $\alpha$ learned                & 0.061 & 0.950 & 0.933 & 0.950 & 0.950 \\
       & $\sigma_a$ sampled              & 0.060 & 0.925 & 0.942 & 0.975 & 0.950 \\
       & Hierarchical spline prior       & 0.024 & 0.950 & 0.938 & 1.000 & 0.950 \\
       & $N(0,5^2)$ coefficients         & 0.026 & 0.950 & 0.940 & 0.975 & 0.975 \\
       & $N(0,10^2)$ coefficients        & 0.019 & 0.925 & 0.937 & 0.975 & 0.975 \\
       & Combined (Section~\ref{sec:functional-model})            & 0.021 & 0.950 & 0.950 & 0.950 & 0.950 \\
    \addlinespace
    Semi- & Baseline& 0.095 & 0.850 & 0.852 & 0.925 & 0.750 \\
    parametric & $\alpha$ learned        & 0.093 & 0.850 & 0.862 & 0.950 & 0.800 \\
       & $\sigma_a$ sampled              & 0.096 & 0.900 & 0.879 & 0.950 & 0.800 \\
       & Hierarchical spline prior       & 0.070 & 0.925 & 0.867 & 0.950 & 0.775 \\
       & $N(0,5^2)$ coefficients         & 0.054 & 0.950 & 0.863 & 0.900 & 0.800 \\
       & $N(0,10^2)$ coefficients        & 0.043 & 0.950 & 0.858 & 0.875 & 0.775 \\
       & Combined (Section~\ref{sec:functional-model})            & 0.067 & 0.900 & 0.915 & 0.925 & 0.900 \\
    \bottomrule
  \end{tabular}
\end{table}

\section{MCMC settings and diagnostics for the recurrent seizure analysis}
\label{sec:supp-mcmc-diagnostics}

We ran four chains. Chain~1 starts from the fixed-effect estimates and
predicted rat effects of the preliminary negative-binomial mixed model; the
other three chains start from coefficients drawn from a normal distribution
centered at those estimates with four times their estimated covariance, and
from jittered rat effects. The initial CRM weights are obtained by
alternating 25 times between the conditionals \eqref{eq:supp-u-conditional}
of $u$ and \eqref{eq:supp-w-conditional} of $w$ at the starting values of
$(\beta,a)$, keeping the state with the highest log posterior; this places
the total CRM mass in the region supported by the posterior. Each chain ran
for 30,000 iterations; we discarded the first 5,000 iterations of each chain,
during which step sizes were adapted, and retained every fifth draw, giving
20,000 posterior draws. Table~\ref{tab:supp-rhat} 
reports the potential scale reduction factor \citep{gelman1992inference} and
the effective sample size, both computed with the \texttt{coda} package
\citep{plummer2006coda}, for the monitored quantities, including the log
likelihood $\sum_{i=1}^n\log p(y_i\mid\theta_i,w)$ with
$p(y_i\mid\theta_i,w)$ as in \eqref{eq:supp-likelihood}; all potential scale reduction factors are at most
$1.001$. The post-adaptation acceptance rates were $0.52$--$0.61$ for the
$(\beta,a)$ update and $0.54$--$0.55$ for the block CRM updates. Each chain took
about eight minutes.

\begin{figure}[!ht]
  \centering
  \includegraphics[width=0.82\linewidth]{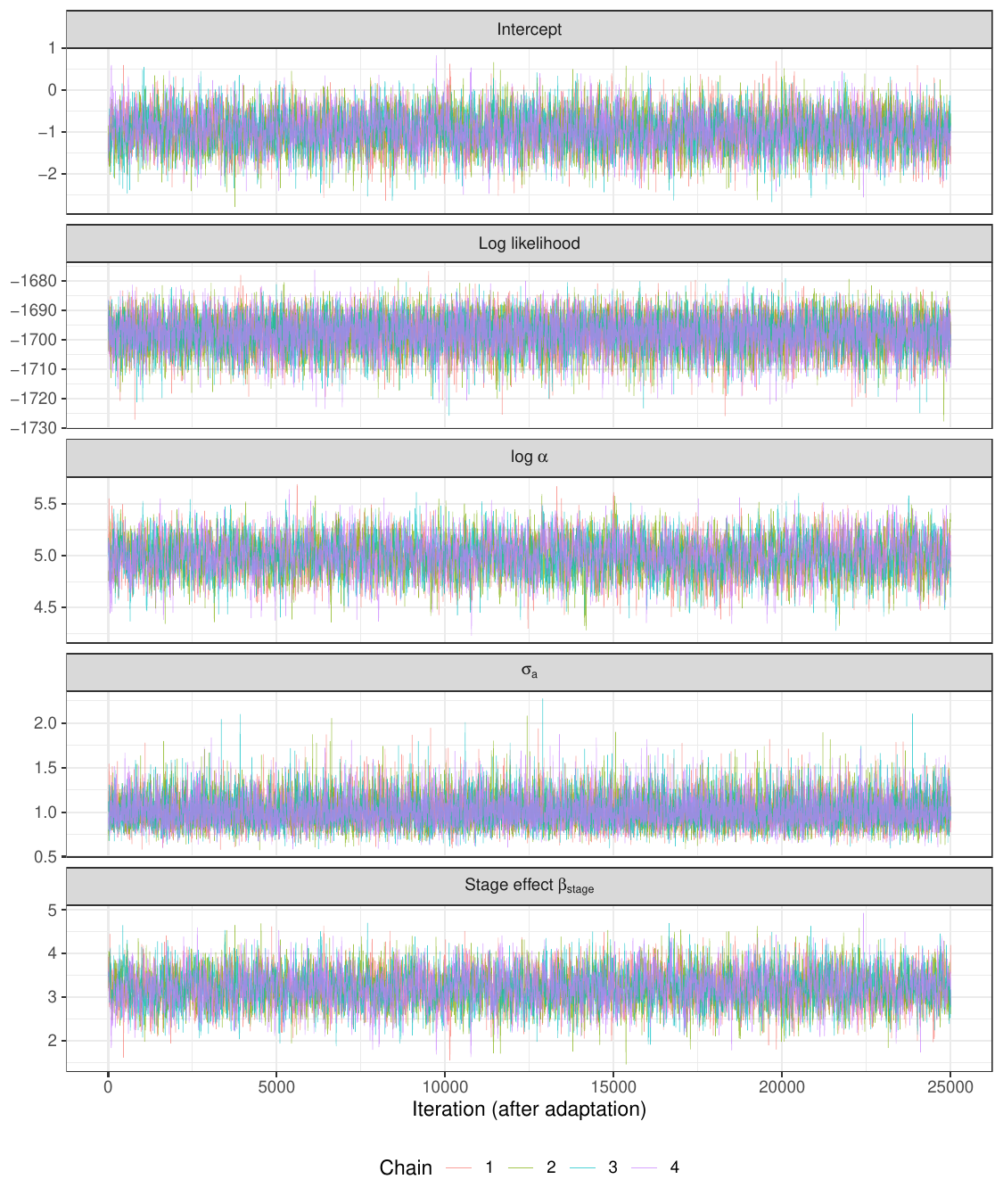}
  \caption{Post-adaptation traces by chain for the intercept, the log
  likelihood, $\log\alpha$, $\sigma_a$ and the stage effect $\beta_{\mathrm{stage}}$ in the
  recurrent seizure analysis (every fifth iteration).}
  \label{fig:supp-trace-parameters}
\end{figure}

\begin{figure}[!ht]
  \centering
  \includegraphics[width=0.82\linewidth]{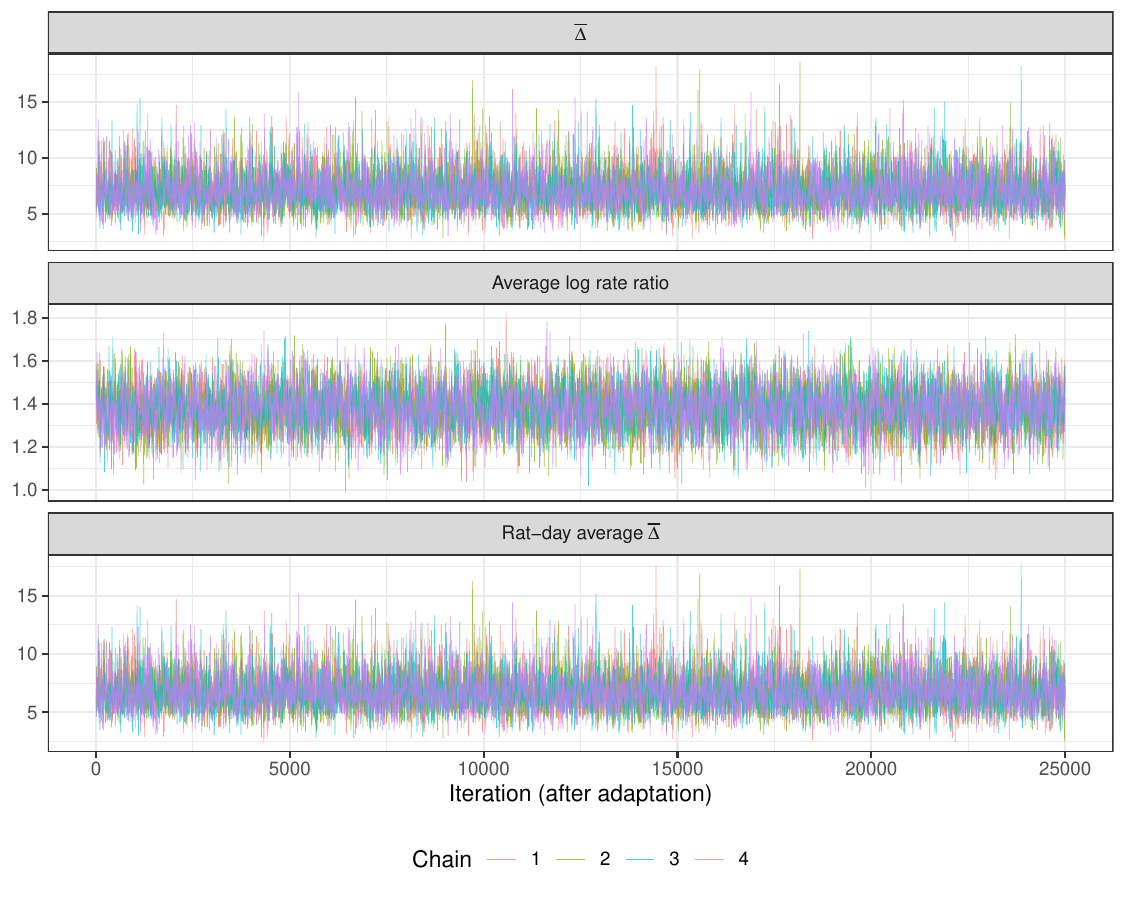}
  \caption{Post-adaptation traces by chain for the average difference
  $\bar\Delta$, the average log rate ratio and the rat-day average difference
  in the recurrent seizure analysis (every fifth iteration).}
  \label{fig:supp-functional-traces}
\end{figure}

\begin{table}[!ht]
\centering
\caption{Convergence diagnostics across four chains for the recurrent seizure
analysis: potential scale reduction factor ($\hat R$) and effective sample size
(ESS) from 20,000 retained draws.}
\label{tab:supp-rhat}
\begin{tabular}{lrr}
\toprule
Quantity & $\hat R$ & ESS \\
\midrule
Intercept & 1.000 & 6844 \\
Stage effect $\beta_{\mathrm{stage}}$& 1.000 & 6022 \\
Stage 1 spline coefficients & 1.000 & 6098--9389 \\
Stage 2 spline coefficients & 1.000 & 4634--8500 \\
$\sigma_a$ & 1.000 & 15384 \\
$\log\alpha$ & 1.001 & 4786 \\
Log likelihood & 1.000 & 13273 \\
Log total CRM mass & 1.001 & 5189 \\
\bottomrule
\end{tabular}
\end{table}
\noindent Figures~\ref{fig:supp-trace-parameters}
and~\ref{fig:supp-functional-traces} show traces by chain for the main
parameters and for the stage contrasts: the average difference
$\bar\Delta=\tfrac{1}{13}\sum_{t=1}^{13}\Delta(t)$ and the average log rate
ratio $\tfrac{1}{13}\sum_{t=1}^{13}\log RR(t)$ between Stage~2 and Stage~1,
and the rat-day average difference of Appendix~\ref{sec:supp-additional-results}. The four chains overlap
throughout, with no evident trends. The effective sample sizes of $\bar\Delta$
and the average log rate ratio are about 10,000 and 8,400, with Monte Carlo
standard errors of $0.017$ seizures per day and $0.001$.

\section{Additional posterior summaries from the recurrent seizure analysis}
\label{sec:supp-additional-results}

Table~\ref{tab:supp-beta} gives posterior summaries of the regression
coefficients. Their values depend on the spline basis; the stage contrasts are
summarized by the functionals in Section~\ref{sec:realdata}.
\begin{table}[!ht]
\centering
\caption{Posterior summaries for the \fcdpglm{} regression coefficients in the
recurrent seizure analysis.}
\label{tab:supp-beta}
\begin{tabular}{lrrr}
\toprule
Parameter & Posterior mean & 2.5\% & 97.5\% \\
\midrule
Intercept & $-0.969$ & $-1.878$ & $-0.066$ \\
Stage effect $\beta_{\mathrm{stage}}$& $3.203$ & $2.392$ & $4.022$ \\
Stage 1 spline 1 & $0.834$ & $0.040$ & $1.625$ \\
Stage 1 spline 2 & $4.392$ & $2.308$ & $6.490$ \\
Stage 1 spline 3 & $-0.606$ & $-2.052$ & $0.836$ \\
Stage 2 spline 1 & $0.075$ & $-0.227$ & $0.409$ \\
Stage 2 spline 2 & $0.006$ & $-0.421$ & $0.538$ \\
Stage 2 spline 3 & $-0.314$ & $-0.783$ & $0.058$ \\
\bottomrule
\end{tabular}
\end{table}
Table~\ref{tab:supp-bydelta} gives the stage-specific mean trajectories and the
stage difference $\Delta(t)$ by day for a typical rat with one day of exposure.
\begin{table}[!ht]
\centering
\caption{Posterior means and $95\%$ credible intervals under \fcdpglm{} of the
stage-specific mean trajectories $m_d(t)$ and the stage difference $\Delta(t)$,
in seizures per day, for a typical rat ($a=0$) with one day of exposure.}
\label{tab:supp-bydelta}
\small
\begin{tabular}{rccc}
\toprule
Day & Stage 1 & Stage 2 & $\Delta(t)$ \\
\midrule
1 & $0.44$ $(0.16,0.97)$ & $9.62$ $(5.73,14.98)$ & $9.18$ $(5.42,14.35)$ \\
2 & $1.03$ $(0.50,1.87)$ & $9.84$ $(5.98,15.16)$ & $8.81$ $(5.30,13.66)$ \\
3 & $2.25$ $(1.26,3.68)$ & $10.07$ $(6.18,15.47)$ & $7.82$ $(4.71,12.20)$ \\
4 & $4.16$ $(2.40,6.70)$ & $10.30$ $(6.31,15.87)$ & $6.14$ $(3.51,9.90)$ \\
5 & $5.87$ $(3.42,9.48)$ & $10.50$ $(6.41,16.21)$ & $4.62$ $(2.15,8.00)$ \\
6 & $6.31$ $(3.74,10.01)$ & $10.66$ $(6.49,16.50)$ & $4.35$ $(1.98,7.66)$ \\
7 & $5.61$ $(3.32,8.81)$ & $10.76$ $(6.59,16.64)$ & $5.15$ $(2.69,8.69)$ \\
8 & $4.50$ $(2.60,7.22)$ & $10.81$ $(6.62,16.65)$ & $6.32$ $(3.57,10.28)$ \\
9 & $3.52$ $(2.01,5.74)$ & $10.81$ $(6.62,16.60)$ & $7.29$ $(4.28,11.55)$ \\
10 & $2.83$ $(1.63,4.57)$ & $10.75$ $(6.60,16.48)$ & $7.91$ $(4.74,12.37)$ \\
11 & $2.36$ $(1.33,3.90)$ & $10.63$ $(6.51,16.28)$ & $8.27$ $(4.95,12.85)$ \\
12 & $2.04$ $(1.01,3.69)$ & $10.46$ $(6.38,16.07)$ & $8.41$ $(4.94,13.15)$ \\
13 & $1.85$ $(0.69,3.96)$ & $10.23$ $(6.23,15.71)$ & $8.38$ $(4.80,13.28)$ \\
\bottomrule
\end{tabular}
\end{table}
Table~\ref{tab:supp-contrasts} compares the stage contrasts of Section~\ref{sec:realdata}
under \textit{fc}DPGLM with those from the Poisson and negative-binomial
GLMMs fitted with the same mean model. For \textit{fc}DPGLM we report
posterior means with $95\%$ credible intervals given by the $2.5\%$ and
$97.5\%$ posterior quantiles. For the GLMMs we drew $2{,}000$ values from the
asymptotic normal distribution of the fixed effects and report the mean and
the $2.5\%$ and $97.5\%$ quantiles of the resulting contrasts.
\begin{table}[!ht]
\centering
\caption{Stage contrasts over the common follow-up period in the recurrent
seizure analysis under \textit{fc}DPGLM and the Poisson and negative-binomial
(NB) GLMMs: estimates and $95\%$ intervals.}
\label{tab:supp-contrasts}
\small
\setlength{\tabcolsep}{5pt}
\begin{tabular}{lccc}
\toprule
Contrast & \textit{fc}DPGLM & Poisson GLMM & NB GLMM \\
\midrule
Average difference $\bar\Delta$ (seizures per day) & $7.13$ $(4.33,11.01)$ & $7.07$ $(4.44,10.65)$ & $7.11$ $(4.29,11.01)$ \\
Average log rate ratio & $1.381$ $(1.178,1.587)$ & $1.334$ $(1.252,1.413)$ & $1.434$ $(1.208,1.659)$ \\
Geometric-mean rate ratio & $4.00$ $(3.25,4.89)$ & $3.80$ $(3.50,4.11)$ & $4.22$ $(3.35,5.25)$ \\
\bottomrule
\end{tabular}
\end{table}
The
posterior median of the concentration parameter was $\alpha=149$ $(102,210)$,
indicating strong shrinkage of the reference distribution toward its
negative-binomial base, with some departures from it (Section~\ref{sec:discussion}). The
spline scales have posterior medians $\tau_1=1.27$ and $\tau_2=0.085$. The
realized standard deviation of the $21$ rat intercepts has posterior mean
$0.973$ $(0.848,1.098)$, close to the posterior median of the random-effect
standard deviation $\sigma_a$. Averaging over the observed rat-days instead of over the days of the common
follow-up period gives the rat-day average difference
\[
  n^{-1}\sum_{i=1}^n\bigl[\exp\{\beta_0+\beta_{\mathrm{stage}}+B(t_i)^\top\gamma_2\}
  -\exp\{\beta_0+B(t_i)^\top\gamma_1\}\bigr],
\]
the average difference between the two stage-specific mean curves, evaluated at
the time of every observed rat-day, for a typical rat ($a=0$) with one day of
exposure. Its posterior mean is $6.77$ $(4.13,10.43)$ seizures per day. Because it
weights days by how often they were observed, it depends on the design.

Table~\ref{tab:supp-dispersion} gives the dispersion index, the zero probability
and the exceedance probability $\Pr(Y\geq20\mid\mu)$ at selected conditional
means, as shown in Figure~\ref{fig:case-dispersion}.
\begin{table}[!ht]
\centering
\caption{Conditional count distribution at selected means $\mu$ in the
recurrent seizure analysis: dispersion index $V(\mu)/\mu$, zero probability
$\Pr(Y=0\mid\mu)$ and exceedance probability $\Pr(Y\geq20\mid\mu)$ under
\textit{fc}DPGLM (posterior mean, with $95\%$ credible interval for the
dispersion index and the exceedance probability) and the negative-binomial (NB)
GLMM (at its estimated size parameter) and Poisson GLMM. The Poisson dispersion index is $1$ and its zero
probability is $e^{-\mu}$.}
\label{tab:supp-dispersion}
\footnotesize
\setlength{\tabcolsep}{4pt}
\begin{tabular}{rccccccc}
\toprule
 & \multicolumn{2}{c}{$V(\mu)/\mu$} & \multicolumn{2}{c}{$\Pr(Y=0\mid\mu)$} &
 \multicolumn{3}{c}{$\Pr(Y\geq20\mid\mu)$} \\
\cmidrule(lr){2-3}\cmidrule(lr){4-5}\cmidrule(lr){6-8}
$\mu$ & \textit{fc}DPGLM & NB & \textit{fc}DPGLM & NB & \textit{fc}DPGLM & NB & Poisson \\
\midrule
$1.0$ & $3.1$ $(2.6,3.5)$ & $2.1$ & $0.59$ & $0.51$ & $0.000$ $(0.000,0.000)$ & $0.000$ & $0.000$ \\
$2.0$ & $4.5$ $(4.0,5.2)$ & $3.1$ & $0.46$ & $0.35$ & $0.001$ $(0.001,0.002)$ & $0.000$ & $0.000$ \\
$5.1$ & $7.5$ $(6.7,8.3)$ & $6.5$ & $0.27$ & $0.18$ & $0.038$ $(0.029,0.047)$ & $0.030$ & $0.000$ \\
$10.1$ & $9.4$ $(8.4,10.4)$ & $11.9$ & $0.15$ & $0.10$ & $0.170$ $(0.152,0.189)$ & $0.154$ & $0.004$ \\
$20.1$ & $9.4$ $(8.0,11.0)$ & $22.6$ & $0.05$ & $0.05$ & $0.488$ $(0.457,0.520)$ & $0.374$ & $0.539$ \\
$30.9$ & $8.4$ $(6.2,11.1)$ & $34.2$ & $0.02$ & $0.04$ & $0.756$ $(0.708,0.801)$ & $0.518$ & $0.985$ \\
\bottomrule
\end{tabular}
\end{table}
Figure~\ref{fig:supp-rat-effects} displays the posterior rat-specific
effects on the multiplicative rate scale.
The conditional stage contrasts reported in Section~\ref{sec:realdata} correspond to a
typical rat with $a=0$. Because the random intercept is Gaussian, integrating
over a new rat multiplies each conditional mean by $\exp(\sigma_a^2/2)$, whose
posterior median is $1.63$ ($95\%$ credible interval $1.30$--$2.77$). Its
posterior mean is not finite, because the half-Cauchy prior gives $\sigma_a$
a polynomial posterior tail, so we report quantiles. Thus, population-average
absolute count differences are about $1.6$ times the corresponding
conditional differences, whereas rate ratios are unchanged.

\section{Additional details and results for the simulation study}
\label{sec:supp-simulation}

\subsection{MCMC settings and computation of the functionals}
\label{sec:supp-sim-mcmc}

For each of the $1{,}600$ simulated data sets (eight settings with $200$
replicates each; Section~\ref{sec:sim-design}), we fit \fcdpglm\ using
the posterior simulation described in Appendix~\ref{sec:supp-mcmc}, with
one chain of $3{,}000$ iterations started at the true parameter values; the first $1{,}000$
iterations were used for adaptation and discarded. Across the eight settings,
the median effective sample size of the $2{,}000$ retained draws was
$190$--$354$ for the average log rate ratio and $70$--$113$ for $\log\alpha$.
The posterior median of $\alpha$ was about $155$ ($R=10$) and $250$ ($R=21$)
under the NB scenario and about $93$--$108$ under the semiparametric scenario.

The distributional functionals (the dispersion index, the zero probability and
the exceedance probabilities) are computed for each data set from $1{,}000$
posterior draws (every second retained post-adaptation draw), and for the mixed models from $1{,}000$ draws
of the fixed effects (and, for the exceedance probability at a given mean, of
the logarithm of the size parameter). For the functionals evaluated for a
typical rat at each day, a posterior draw can occasionally place the
typical-rat curve above the truncation point $K$ at some day, where the mean
constraint has no solution, because the sampler enforces $0<\mu_i<K$ only at
the observed rat-days. This happened in $1.3\%$ of the data sets (all with ten rats, where $K$ is
smaller), for one or two of the $1{,}000$ draws;
such draws are left out of the summaries of the affected quantities.

\subsection{Pointwise coverage}
\label{sec:supp-sim-coverage}

\begin{figure}[!ht]
  \centering
  \includegraphics[width=\linewidth]{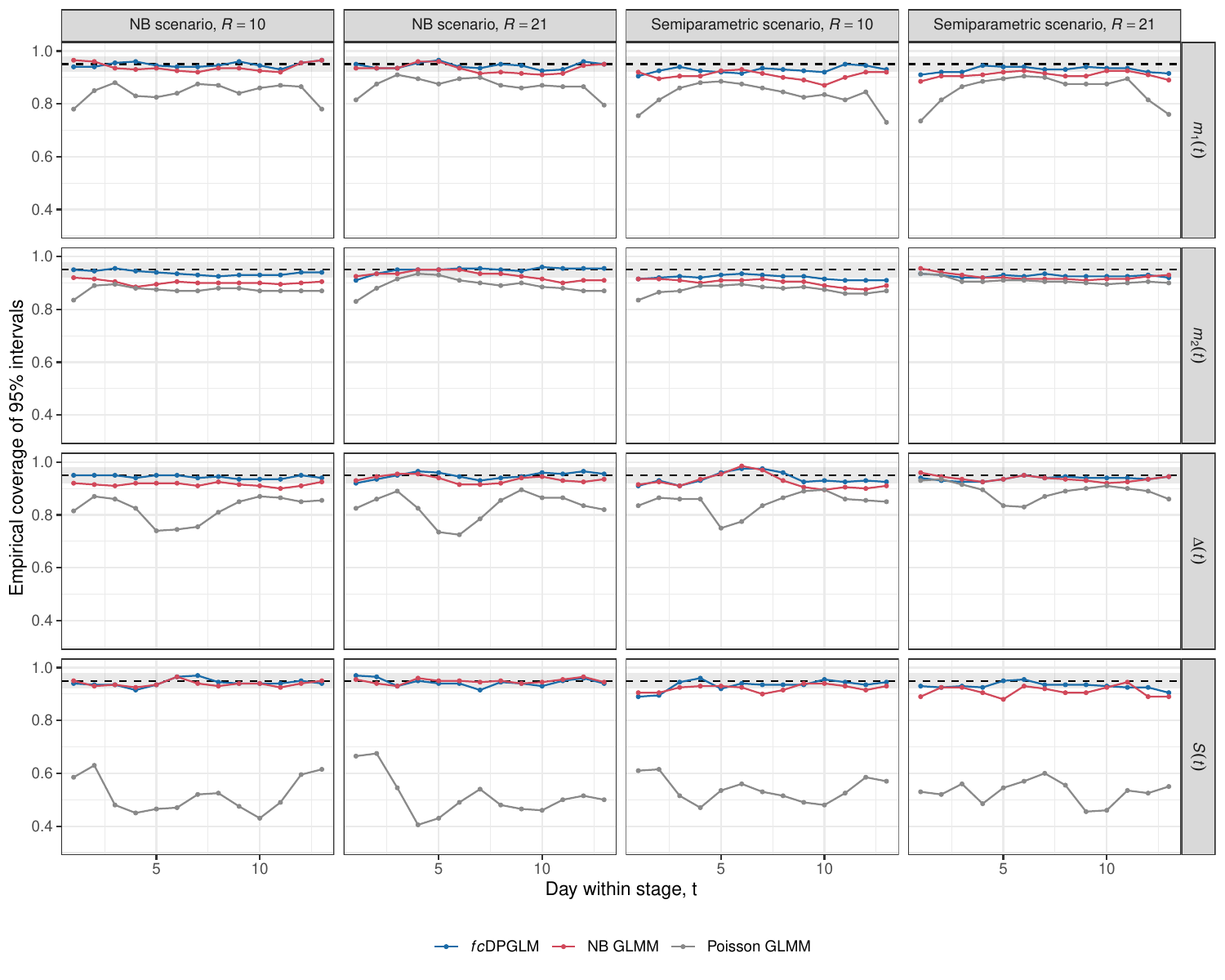}
  \caption{Pointwise coverage of $95\%$ intervals by day. Grey band:
  $0.95\pm1.96$ Monte Carlo standard errors, about $0.92$--$0.98$ with $200$
  replicates.}
  \label{fig:supp-sim-coverage}
\end{figure}

Figure~\ref{fig:supp-sim-coverage} shows the pointwise coverage of the
$95\%$ intervals for $m_1(t)$, $m_2(t)$, $\Delta(t)$ and $S(t)$. \fcdpglm\
stays within or close to the Monte Carlo band around $0.95$ on every day and in
every setting. The NB mixed model undercovers slightly, mainly for $m_2(t)$ and
$\Delta(t)$ with ten rats and for $m_1(t)$ and $S(t)$ under the semiparametric
scenario, and the Poisson mixed model undercovers throughout.

\subsection{Recovery of the distributional curves}
\label{sec:supp-sim-curves}
\begin{figure}[!ht]
  \centering
  \includegraphics[width=\linewidth]{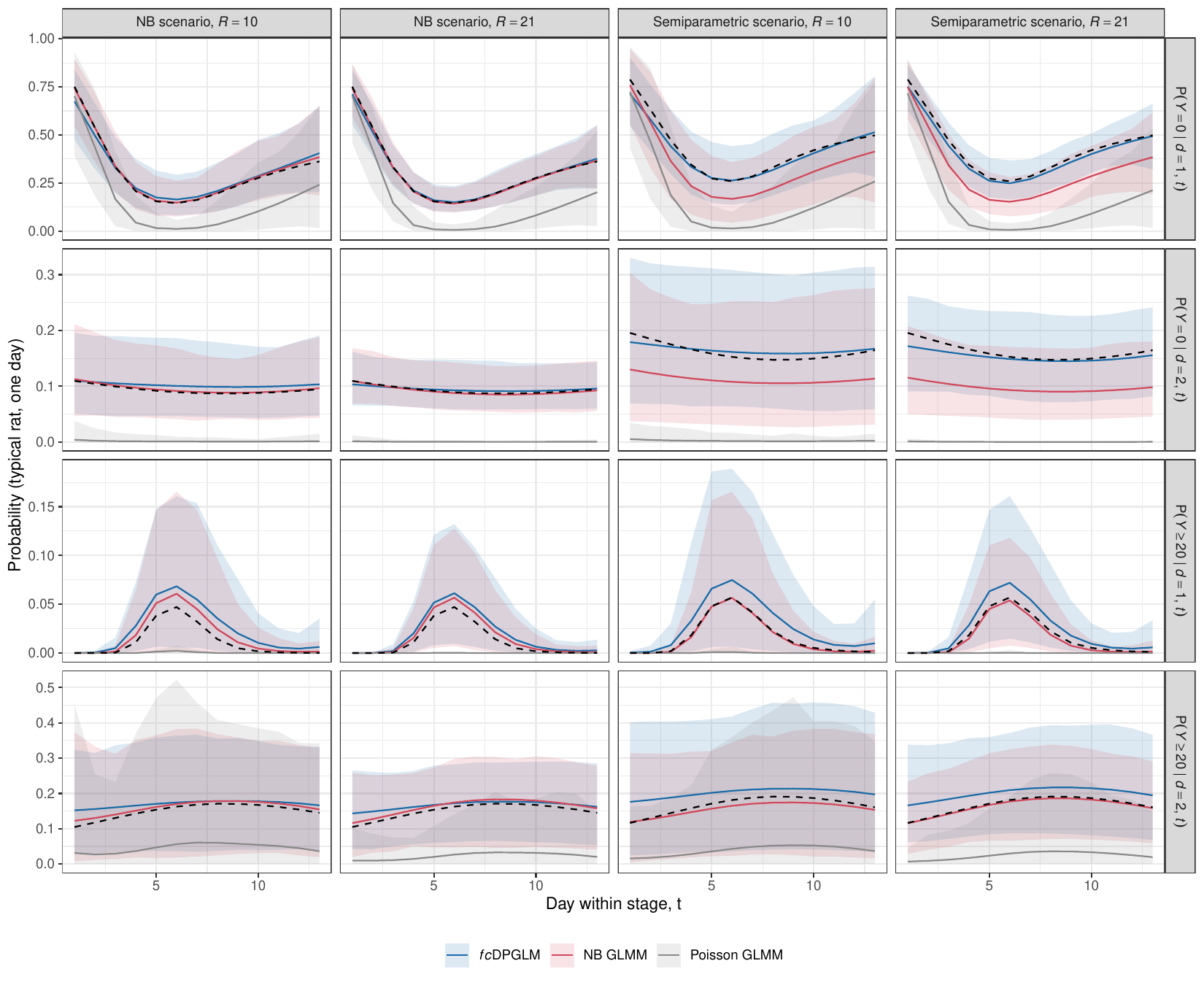}
  \caption{True (dashed) and estimated $\Pr(Y=0\mid d,t)$ and
  $\Pr(Y\geq20\mid d,t)$ for a typical rat with one day of exposure: mean and $2.5$--$97.5\%$ range of
  the point estimates over $200$ replicates.}
  \label{fig:supp-sim-distcurves}
\end{figure}

Figure~\ref{fig:supp-sim-distcurves} compares the true stage- and
day-specific probabilities $\Pr(Y=0\mid d,t)$ and $\Pr(Y\geq20\mid d,t)$ for a
typical rat with one day of exposure with the mean and $2.5$--$97.5\%$ range
of the point estimates over the $200$ replicates.

\subsection{Exceedance probabilities}
\label{sec:supp-sim-exceed}

The distributional estimands at a given mean are evaluated on grids of means:
$\mu=0.5$, $1$, $2$, $5$, $10$, $20$, $30$ for the dispersion index $V(\mu)/\mu$ of
Table~\ref{tab:sim-results}, Panel~C, and $\mu=1$, $2$, $5$, $10$, $15$, $20$, $30$ for the
exceedance probability. Table~\ref{tab:supp-sim-exceed} reports two
exceedance functionals at thresholds $c\in\{10,20,40\}$: the exceedance
probability
$\Pr(Y\geq c\mid\mu)$ of the conditional distribution, integrated over its grid of
means, and the exceedance probability averaged over the
rat-days of each data set, $n^{-1}\sum_i\Pr(Y_i\geq c)$, whose truth uses the
true rat effects. \fcdpglm\ intervals are $95\%$ credible intervals given by
the $2.5\%$ and $97.5\%$ posterior quantiles over $1{,}000$ posterior draws (every second retained post-adaptation draw), as for the other
distributional functionals. NB mixed-model intervals use $1{,}000$ draws of
the fixed effects and of the logarithm of the size parameter from their
asymptotic normal distribution, with rat effects at their predictions.
Under the NB scenario the correctly specified NB mixed model is more accurate
for the exceedance probability given the mean, although both integrated RMSEs
are at most about $0.02$; the two methods are about equally accurate for the
average over rat-days, and both are close to nominal coverage given the mean.
Under the semiparametric scenario \fcdpglm\ is more
accurate for every threshold and both functionals, and the NB intervals given
the mean almost never cover. \fcdpglm\ undercovers mainly at $c=10$ under the
semiparametric scenario, where a bias of about $0.02$ gives coverage of
$0.64$--$0.81$, and less so at $c=40$ ($0.85$--$0.93$). Results under the null-shape mean
curves are nearly identical and are omitted.
\begin{table}[!ht]
  \centering
  \caption{Exceedance probabilities: integrated RMSE and mean coverage of
  $95\%$ intervals for $\Pr(Y\geq c\mid\mu)$ over the grid of means, and RMSE and
  coverage for the exceedance probability averaged over the rat-days, by
  threshold $c$. $200$ replicates per setting (shape-difference mean curves).}
  \label{tab:supp-sim-exceed}
  \small
  \begin{tabular}{llrrrrrrrr}
    \toprule
    & & \multicolumn{4}{c}{$\Pr(Y\geq c\mid\mu)$} & \multicolumn{4}{c}{Averaged over rat-days} \\
    \cmidrule(lr){3-6}\cmidrule(lr){7-10}
    & & \multicolumn{2}{c}{\fcdpglm} & \multicolumn{2}{c}{NB GLMM} & \multicolumn{2}{c}{\fcdpglm} & \multicolumn{2}{c}{NB GLMM} \\
    \cmidrule(lr){3-4}\cmidrule(lr){5-6}\cmidrule(lr){7-8}\cmidrule(lr){9-10}
    Scenario & $c$ & IRMSE & Cov & IRMSE & Cov & RMSE & Cov & RMSE & Cov \\
    \midrule
    \multicolumn{10}{l}{$R=21$ rats} \\
    NB & 10 & 0.015 & 0.95 & 0.009 & 0.96 & 0.014 & 0.95 & 0.013 & 1.00 \\
     & 20 & 0.014 & 0.96 & 0.004 & 0.96 & 0.011 & 0.96 & 0.011 & 1.00 \\
     & 40 & 0.009 & 0.95 & 0.002 & 0.96 & 0.008 & 0.94 & 0.007 & 1.00 \\
    \addlinespace
    Semiparametric & 10 & 0.023 & 0.64 & 0.113 & 0.00 & 0.024 & 0.69 & 0.076 & 0.80 \\
     & 20 & 0.013 & 0.91 & 0.113 & 0.04 & 0.010 & 0.97 & 0.060 & 0.92 \\
     & 40 & 0.014 & 0.85 & 0.039 & 0.10 & 0.009 & 0.90 & 0.017 & 0.98 \\
    \midrule
    \multicolumn{10}{l}{$R=10$ rats} \\
    NB & 10 & 0.022 & 0.96 & 0.013 & 0.94 & 0.019 & 0.95 & 0.018 & 1.00 \\
     & 20 & 0.021 & 0.96 & 0.007 & 0.94 & 0.015 & 0.93 & 0.015 & 1.00 \\
     & 40 & 0.016 & 0.93 & 0.003 & 0.95 & 0.010 & 0.96 & 0.010 & 1.00 \\
    \addlinespace
    Semiparametric & 10 & 0.030 & 0.72 & 0.117 & 0.01 & 0.028 & 0.81 & 0.074 & 0.96 \\
     & 20 & 0.020 & 0.90 & 0.114 & 0.10 & 0.015 & 0.96 & 0.057 & 0.98 \\
     & 40 & 0.021 & 0.86 & 0.039 & 0.17 & 0.010 & 0.93 & 0.018 & 0.98 \\
    \bottomrule
  \end{tabular}
\end{table}

\clearpage
\subsection{Results for reduced sample size}
\label{sec:supp-sim-dist10}

\begin{figure}[!ht]
  \centering
  \includegraphics[width=\linewidth]{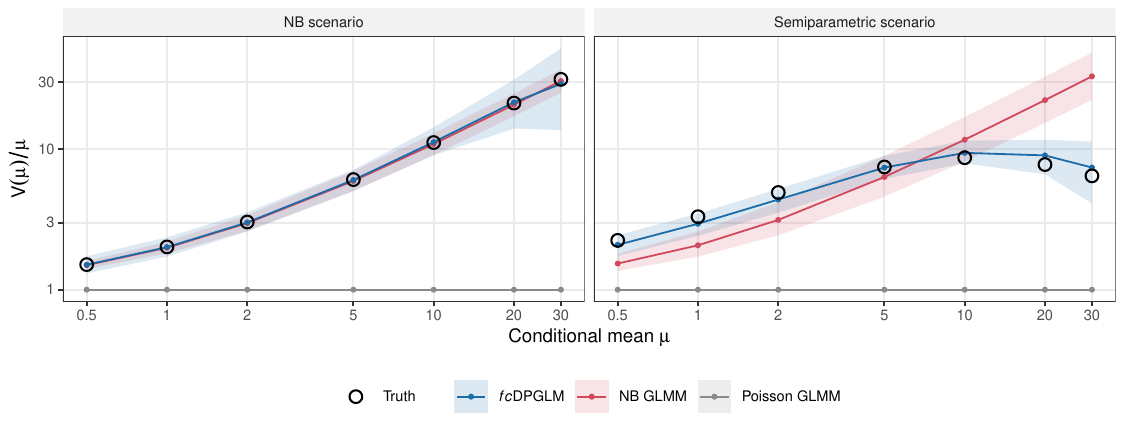}
  \caption{As Table~\ref{tab:sim-results}, Panel~C, for $R=10$ rats. Open circles: truth. Solid lines and bands: mean and $2.5$--$97.5\%$ range of the estimates over $200$ replicates.}
  \label{fig:supp-sim-disp10}
\end{figure}

Tables~\ref{tab:supp-sim-mean10} and~\ref{tab:supp-sim-dist10} and Figure~\ref{fig:supp-sim-disp10} repeat Panels~A, B and~C of Table~\ref{tab:sim-results} for $R=10$
rats. Errors are larger with fewer rats, but the comparisons between methods
are unchanged.

\begin{table}[!ht]
  \centering
  \caption{As Table~\ref{tab:sim-results}, Panel~A, for $R=10$ rats.}
  \label{tab:supp-sim-mean10}
  \small
  \resizebox{\linewidth}{!}{%
  \begin{tabular}{llrrrrrrrrrr}
    \toprule
    & & \multicolumn{3}{c}{Avg.\ log RR} &
      \multicolumn{2}{c}{$\bar\Delta$}
      & \multicolumn{3}{c}{$\Delta(t)$} & \multicolumn{2}{c}{$S(t)$} \\
    \cmidrule(lr){3-5}\cmidrule(lr){6-7}\cmidrule(lr){8-10}\cmidrule(lr){11-12}
    Scenario & Method & Bias & RMSE & Cov & RMSE & Cov & IRMSE &
      Cov & Sim & IRMSE & Sim \\
    \midrule
    NB & \fcdpglm & $-0.018$ & 0.143 & 0.965 & 2.44 &
      0.935 & 2.71 & 0.944 & 0.930 & 0.265 & 0.925 \\
     & NB GLMM & $0.012$ & 0.145 & 0.965 & 2.51 &
      0.895 & 2.98 & 0.915 & 0.895 & 0.277 & 0.950 \\
     & Poisson GLMM & $0.033$ & 0.165 & 0.555 & 2.46 &
      0.885 & 2.99 & 0.824 & 0.620 & 0.347 & 0.210 \\
    \addlinespace
    Semiparametric & \fcdpglm & $0.000$ & 0.163 & 0.965 & 2.47 &
      0.925 & 2.68 & 0.937 & 0.920 & 0.316 & 0.920 \\
     & NB GLMM & $0.054$ & 0.185 & 0.930 & 2.55 &
      0.890 & 2.91 & 0.926 & 0.925 & 0.331 & 0.910 \\
     & Poisson GLMM & $0.050$ & 0.181 & 0.510 & 2.47 &
      0.865 & 2.73 & 0.846 & 0.680 & 0.341 & 0.260 \\
    \bottomrule
  \end{tabular}}
\end{table}

\begin{table}[!ht]
  \centering
  \caption{As Table~\ref{tab:sim-results}, Panel~B, for $R=10$ rats.}
  \label{tab:supp-sim-dist10}
  \small
  \resizebox{\linewidth}{!}{%
  \begin{tabular}{llrrrrrrrr}
    \toprule
    & & \multicolumn{3}{c}{Bias, averaged over rat-days}
      & \multicolumn{2}{c}{$\Pr(Y=0\mid d,t)$} & \multicolumn{2}{c}{$\Pr(Y\geq20\mid\mu)$} & Excess \\
    \cmidrule(lr){3-5}\cmidrule(lr){6-7}\cmidrule(lr){8-9}\cmidrule(lr){10-10}
    Scenario & Method & $\Pr(Y=0)$ & Variance & $\Pr(Y\geq20)$ & IRMSE & Cov & IRMSE & Cov & RPS \\
    \midrule
    NB & & (0.229) & (208) & (0.152) & & & & & \\
     & \fcdpglm & $-0.002$ & $2.4$ & $-0.002$ &
      0.065 & 0.95 & 0.021 & 0.96 & 0.146 \\
     & NB GLMM & $-0.001$ & $-12.4$ & $-0.007$ &
      0.061 & 0.90 & 0.007 & 0.94 & 0.145 \\
     & Poisson GLMM & $-0.101$ & $-198$ & $0.015$ &
      0.141 & 0.25 & 0.200 & -- & 1.477 \\
    \addlinespace
    Semiparametric & & (0.308) & (73.4) & (0.206) & & & & & \\
     & \fcdpglm & $-0.015$ & $5.8$ & $-0.010$ &
      0.082 & 0.91 & 0.020 & 0.90 & 0.153 \\
     & NB GLMM & $-0.060$ & $125$ & $-0.060$ &
      0.117 & 0.60 & 0.114 & 0.10 & 0.441 \\
     & Poisson GLMM & $-0.170$ & $-63.6$ & $-0.033$ &
      0.231 & 0.15 & 0.126 & -- & 0.871 \\
    \bottomrule
  \end{tabular}}
\end{table}

\section{Additional posterior predictive checks for the recurrent seizure analysis}
\label{sec:supp-ppc}

\begin{figure}[!ht]
  \centering
  \includegraphics[width=0.90\linewidth]{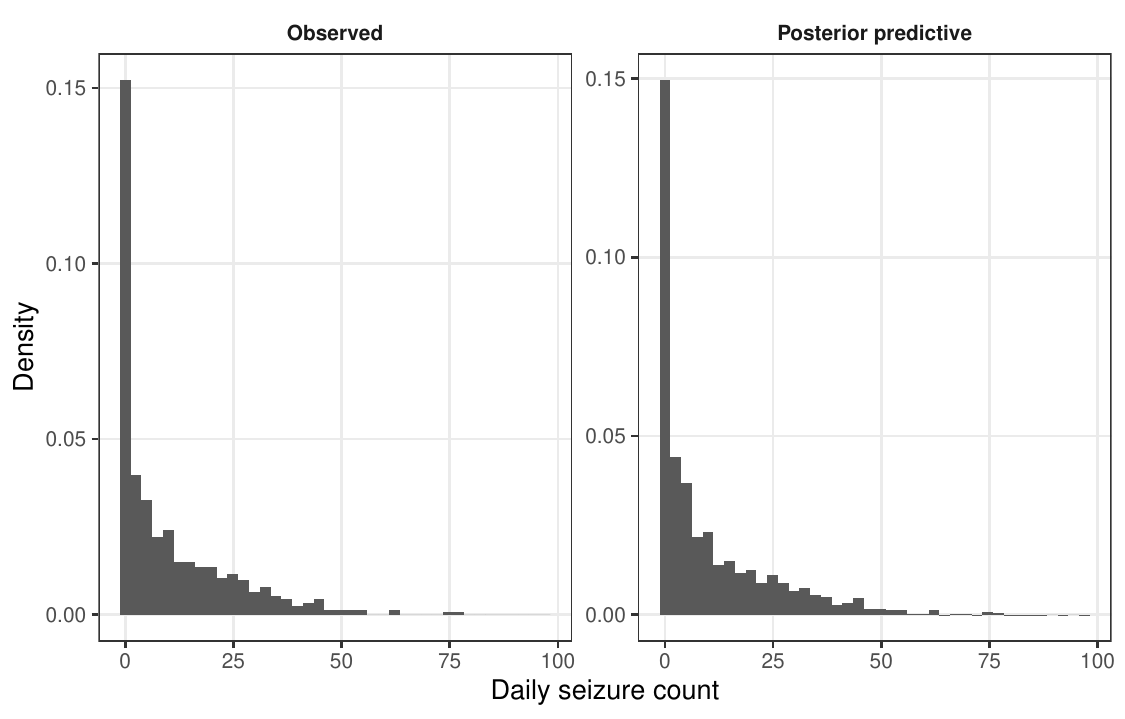}
  \caption{Observed daily-count distribution and posterior predictive
  distribution under \fcdpglm{}.}
  \label{fig:supp-ppc}
\end{figure}

Table~\ref{tab:supp-ppc} compares the observed value of each summary statistic
of the counts with the predictive median and $95\%$ predictive interval under
\textit{fc}DPGLM and the Poisson and negative-binomial GLMMs; these are the
quantities summarized in Figure~\ref{fig:case-comparison}C.
\begin{table}[!ht]
\centering
\caption{Observed value and predictive median ($95\%$ predictive interval) of
summary statistics of the daily seizure counts under \textit{fc}DPGLM and the
Poisson and negative-binomial (NB) GLMMs in the recurrent seizure analysis.}
\label{tab:supp-ppc}
\small
\begin{tabular}{lrccc}
\toprule
Statistic & Observed & \textit{fc}DPGLM & Poisson GLMM & NB GLMM \\
\midrule
Mean & $10.2$ & $10.2$ $(9.3,11.2)$ & $10.2$ $(9.9,10.4)$ & $10.2$ $(9.1,11.4)$ \\
Variance & $175$ & $183$ $(148,221)$ & $105$ $(98,112)$ & $306$ $(211,445)$ \\
Zero proportion & $0.29$ & $0.28$ $(0.24,0.32)$ & $0.11$ $(0.10,0.13)$ & $0.23$ $(0.20,0.26)$ \\
$90$th percentile & $29.3$ & $31$ $(27,33.9)$ & $26$ $(25,27.3)$ & $28$ $(24,33)$ \\
$95$th percentile & $38$ & $38.2$ $(34.2,44)$ & $32$ $(30,33)$ & $43.2$ $(36.2,51.2)$ \\
Maximum & $78$ & $78$ $(56,94)$ & $47$ $(42,56)$ & $150$ $(100,262)$ \\
\bottomrule
\end{tabular}
\end{table}
Table~\ref{tab:supp-ppc-stage} gives stage-specific posterior predictive
summaries under \fcdpglm{}. The fitted means are close to the observed means,
and the predictive variances are close to the observed variances, except that
the Stage~1 predictive variance is somewhat smaller ($58.5$ against $68.4$).
\begin{table}[!ht]
\centering
\caption{Stage-specific observed and posterior predictive summaries under
\fcdpglm{} for the recurrent seizure analysis.}
\label{tab:supp-ppc-stage}
\begin{tabular}{lrrrr}
\toprule
Stage & Observed mean & Fitted mean & Observed variance & Predictive variance \\
\midrule
Stage 1 & 4.486  & 4.590  & 68.390  & 58.508 \\
Stage 2 & 14.315 & 14.270 & 211.283 & 229.846 \\
\bottomrule
\end{tabular}
\end{table}
Figure~\ref{fig:supp-ppc} compares the observed daily-count distribution
with posterior predictive replicates. Together with Figure~\ref{fig:case-comparison}C, these summaries show that \fcdpglm{} reproduces the observed count
distribution: every observed summary in Table~\ref{tab:supp-ppc} lies
within its $95\%$ posterior predictive interval under \fcdpglm{}.

\end{document}